\documentclass[aps,prb,floats,twocolumn,eqsecnum,secnumarabic,byrevtex,superscriptaddress]{revtex4}
\usepackage{graphicx}

\begin{document}

\title{Fock--Bargmann space and classic trajectories}
\author{Gennady F. Filippov}
\author{Sergei Korennov}
\altaffiliation{Currently at: Graduate School of Science, Hokkaido University, Sapporo, Japan} 
\email[E-mail:]{korennov@nucl.sci.hokudai.ac.jp}
\affiliation{Bogoliubov Institute for Theoretical Physics, Kiev, Ukraine}
\author{Arina Sytcheva}
\affiliation{National Shevchenko University, Kiev, Ukraine}
\author{Kiyoshi Kat\=o}
\affiliation{Graduate School of Science, Hokkaido University, Sapporo, Japan}
\date{\today}

\begin{abstract}

In recent years, a new approach to the theory of nuclear reactions leading to a 
break-down
of the interacting subsytems into various channels has been developed. This approach was named 
the Antisymmetrized Molecular Dynamics (AMD), and its main idea lies in the description of the 
nucleons by wave packets in which the antisymmetrization effects (but not other quantum effects) are accounted for. In the present review, the basic principles of AMD are illustrated on the  examples of simplest nuclear systems, and the results are compared with those provided by an exact quantum-mechanical description in the Fock--Bargmann space. The applicability region of AMD is discussed, in particular, in the cases of systems with discrete spectrum, and a relation between the classical AMD trajectories and the quantum distributions is established. At the same time, a new interpretation of Brink orbitals and Slater determinants built on them as eigenfunctions of the coordinate operator defined in the Fock--Bargmann space is proposed. It is shown that these functions form the cluster geometry of nuclear systems.

\end{abstract}

\maketitle

\begin{widetext}

%\tableofcontents

\vspace{1cm}

\end{widetext}

\vspace{1cm}

\section{Introduction}

High-energy heavy-ion collisions lead to a break-down of nuclei and  
scattering of nuclear 
fragments \cite{exp1}. This process appears to be essentially non-equilibrated, with a large number of nucleons participating in it. Therefore, a necessity to apply the kinetic theory to the heavy-ion collision phenomena has been realized some time ago 
\cite{theor1}. However, whereas the standard statistical approach is based on classical 
mechanics, the kinetic theory of nuclear collisions should be based on the quantum equations, or rather such a classical limit of these equations which would reproduce at least the most important quantum features of the nuclear dynamics. Thus, a version of the kinetic approach, 
taking into account the quantum phenomenon of antisymmetrization and its influence on the nucleonic classical trajectories was proposed \cite{Hor1} and was named the Antisymmetrized Molecular Dynamic (AMD).    

Later \cite{Fil1}, it became clear that the nucleonic trajectories, which are the object of study in AMD, are classical limit of wave functions defined in the Hilbert space of integral analytical functions (the Fock--Bargmann space)\cite{F-B}. A peculiar property of these functions is that their variables are both coordinates and momenta, i.e. they are defined in 
the phase space. We recall that, usually, transition from the coordinate to the phase space is done via the Wigner density matrix \cite{Feyn-Wign}, but then one has to deal with the alternating signature of this matrix, which contradicts its physical sense. On the other hand, the density matrix in the Fock--Bargmann space built with the account of the Bargmann measure is positively defined at all values of coordinates and momenta, which is its advantage. Thus the use of the Fock--Bargmann space in order to relate the quantum and classical statistics is well 
justified. In this space, the classical and quantum mechanics are nicely related, including the possibility of the evaluation of the classical results in the limiting quasi-classical region. This is why we pay a special attention here to basic facts concerning the Fock--Bargmann space and illustrate them with examples of exactly solvable problems.  

In Section 2, the definition of the Fock--Bargmann space is given, and the relation between the wave functions defined in the 
coordinate (or momentum) representation and their images in the Fock--Bargmann space is shown.
Section 3 is devoted to the discussion on phase trajectories of those systems wave functions of which were defined in the 
previous Section. Influence of the interaction between the nucleons on their wave functions and phase trajectories is studied 
in Section 4. In Section 5, the breathing of wave packets usually used in AMD is introduced in the equations of their motion. The conclusions are presented in the final Section.

\section{Fock--Bargmann space}

Although the features of the Fock--Bargmann space and functions defined in it are not 
widely known, some of the constructions which are often used in theoretical nuclear physics are 
directly related to this space.

\subsection{The Brink--Bloch orbital}

Among the simplest examples of the wave functions defined in the Fock--Bargmann space, one 
will be specially discussed here; it is the Brink--Bloch orbital, 
\begin{eqnarray}
\label{a1}
\phi_{\bf r}({\bf R})=\pi^{-3/4}\exp\{-{{\bf r}^2\over2}+\sqrt{2}({\bf
Rr})-
{{\bf R}^2\over2}\},
\end{eqnarray}
which was originally proposed \cite{Br} for the analysis of the cluster structure of light nuclei. Here, ${\bf r}$ is the three-dimensional coordinate vector of a nucleon, and ${\bf R}$
is a vector parameter, introduced to minimize the energy of the nucleus, after a trial functions is constructed from (\ref{a1}). The Brink--Bloch orbital is a generalization of the $s$-orbital 
of a nucleon in the harmonic-oscillator (harmonic-oscillator) field. As the units of length, mass and action we have chosen the oscillator length, the nucleon mass and the Planck constant $\hbar$. 

The complex vector ${\bf R}$, in general, consists of the real and imaginary parts,
\begin{eqnarray}
{\bf R}={\vec{\xi}+i\vec{\eta}\over\sqrt{2}},~~
{\bf S}={\bf R}^*={\vec{\xi}-i\vec{\eta}\over\sqrt{2}},
\end{eqnarray}
where $i$ is the imaginary unit, $\vec{\xi}$ is the radius-vector, and $\vec{\eta}$ is the momentum. We use the Brink's notation ${\bf S}={\bf R}^*$ here.  

The orbital (\ref{a1}) is the integral analytical function of three complex projections $R_x,R_y,R_z$ of the vector ${\bf R}$. Therefore, it may be considered as a wave function of these variables in the (Hilbert) Fock--Bargmann space. Then, two questions are to be answered,
"how it is normalized?" and "isn't it an eigenfunction of an operator in this space?" The first question is solved by the following definition,     

$$I({\bf r},{\bf r}')=\int\phi^*_{{\bf r}'}({\bf R})\phi_{\bf r}({\bf R})
\exp\{-({\bf RS})\}{d\vec{\xi}d\vec{\eta}\over(2\pi)^3}$$
where the improper integration is over all phase space, and
$$\exp\{-({\bf RS})\}{d\vec{\xi}d\vec{\eta}\over(2\pi)^3}$$
is the Bargmann measure\cite{Barg}. The factor $(2\pi)^3$ in the denominator is due to 
the number of all quantum states in the phase space. It is easy to see that
\begin{eqnarray}
I({\bf r},{\bf r}')=\delta({\bf r}-{\bf r}').
\end{eqnarray}
Thus, the functions $\phi_{\bf r}({\bf R})$ with different values of ${\bf r}$  are orthogonal,
and the components of the vector ${\bf r}$ are eigenvalues of these functions. The explicit form of the operator $\hat{\bf r}$ follows the expression 
(\ref{a1}), 
\begin{eqnarray}
\label{a2}
\hat{\bf r}={1\over\sqrt{2}}({\bf R}+\nabla_{\bf R}).
\end{eqnarray}
The relation
\begin{eqnarray}
\label{a3}
\hat{\bf r}\phi_{\bf r}({\bf R})={\bf r}\phi_{\bf r}({\bf R})
\end{eqnarray}
is now evident. As it should be, the spectrum of the operator $\hat{\bf r}$ is continuous only, 
and it is due to this fact that all its eigenvectors are normalized to the $\delta$-function.

In addition to all said above, there is one more point outlining the special role of the Brink--Bloch orbital in the Fock--Bargmann space. It is also the kernel of the integral transformation between a wave function $\Psi({\bf r})$ of the coordinate space and a corresponding wave function $\Phi({\bf R})$ of the Fock--Bargmann space,  
\begin{eqnarray}
\label{a4}
\Phi({\bf R})=\int\phi_{\bf r}({\bf R})\Psi({\bf r})d{\bf r}.
\end{eqnarray}
In many cases the integration (\ref{a4}) can be done analytically.

\subsection{Overlap integral and density matrix}

After the integration of the eigenfunctions
$$\phi^*_{\bf r}({\bf R})\phi_{\bf r}({\bf R})$$
over all their eigenvalues, we arrive to an important expression,
\begin{eqnarray}
\label{a5}
\int\phi^*_{\bf r}({\bf R})\phi_{\bf r}({\bf R})d{\bf r}=\exp({\bf RS}),
\end{eqnarray}
which is often called "the overlap integral" and which, also, serves as the density matrix 
in the Fock--Bargmann space. Here we mean the density matrix of pure states, and we want 
to convince ourselves that the expression (\ref{a5}) satisfy all requirements to be a density 
matrix.

The diagonal elements of the standard single-particle density matrix defined in the coordinate 
space (when its two vector arguments are considered identical) yield expressions for the 
probability distribution of the coordinates \cite{Landau1}. The density matrix (\ref{a5}), too, 
defines the probablility distribution, but in the phase space $\{\vec{\xi},\vec{\eta}\}$ and 
not before being multiplied by the Bargmann measure. The resultant expression for the probability distribution is
\begin{eqnarray}
\label{d1}
\exp({\bf RS})\cdot
\exp\{-({\bf RS})\}{d\vec{\xi}d\vec{\eta}\over(2\pi)^3}=
{d\vec{\xi}d\vec{\eta}\over(2\pi)^3},
\end{eqnarray}
and we arrive to the uniform distribution over the coordinates and momenta with the constant 
density equal to unity. Later on, in the discussion on the multi-particle wave functions we 
shall show how this distribution is affected by the antisymmetrization.

Of interest is the probability distribution for the state 
(\ref{a1}),
\begin{eqnarray}
\label{b1}
\phi^*_{\bf r}({\bf R})\phi_{\bf r}({\bf R})
\exp\{-({\bf RS})\}=
\pi^{3/2}\exp\{-({\bf r}-\vec{\xi})^2\}. \nonumber \\
\end{eqnarray}
This distribution is uniform over the momenta and Gaussian over the coordinates, centered 
at ${\bf r}$. When the orbital (\ref{a1}) is used as a trial function in variational calculations, $\vec{\xi}$ becomes the variational parameter, Eq. (\ref{b1}) is then treated as
the probability distribution over ${\bf r}$, and the variational calculation is reduced to a
search for an optimal location of the center $\vec{\xi}$ of the Gaussian distribution. 

In the classical limit, when $|{\bf r}|\gg 1$, i.e. $|{\bf r}|$ is much greater than the oscillator length, the distribution (\ref{b1}) is reduced to $\delta({\bf r}-\vec{\xi})$, 
as it should be for a distribution in the state with a given value of the radius-vector.

The relation (\ref{a5}) between the density matrix and its expansion over a basis of orthonormalized states illustrates the well-known fact \cite{Landau2} of diagonality of the 
density matrix in the energy representation. The expansion of the density matrix over the 
eigenstates of a Hamiltonian remains diagonal for any Hamiltonian. 

For instance, the density matrix can be expanded over the eigenstates of the Hamiltonian of 
free motion, i.e. over the eigenstates of the momentum operator  
\begin{eqnarray}
\label{c1}
\hat{\bf k}= -{i\over\sqrt{2}}({\bf R}-\nabla_{\bf R}).
\end{eqnarray}
The orthonormalized eigenfunctions of this operator, i.e. plane waves, labeled by the momentum $k$, take the form
\begin{eqnarray}
\label{a6}
\phi_{\bf k}({\bf R})=\pi^{-3/4}\exp\{-{{\bf k}^2\over2}-i\sqrt{2}({\bf
Rk})+
{{\bf R}^2\over2}\},
\end{eqnarray}
which can be proved by solving a first-order differential equation with the explicit form of 
the operator $\hat{\bf k}$. Another way to get $\phi_{\bf k}({\bf R})$ is to perform the 
integration (\ref{a4}) with the plane wave $$\Psi({\bf r})=(2\pi)^{3/2}\exp\{-i({\bf kr})\},$$
defined in the coordinate space.
\begin{eqnarray}
\phi_{\bf k}({\bf R})=
\int\phi_{\bf r}({\bf R})(2\pi)^{3/2}\exp\{-i({\bf kr})\}d{\bf r}.
\end{eqnarray}

The expansion of the overlap integral over the states with definite values of momentum is 
similar to (\ref{a5}),
\begin{eqnarray}
\label{a7}
\exp({\bf RS})=\int\phi^*_{\bf k}({\bf R})\phi_{\bf k}({\bf R})d{\bf k},
\end{eqnarray}
and the probability distribution for the plane wave (\ref{a6}) is
\begin{eqnarray}
\label{b2}
\phi^*_{\bf k}({\bf R})\phi_{\bf k}({\bf R})
\exp\{-({\bf RS})\}=
\pi^{3/2}\exp\{-({\bf k}-\vec{\eta})^2\}. \nonumber \\
\end{eqnarray}
Again, this is a Gaussian dependence, but this time for the momentum. In the phase plane, 
the distribution function reaches its maxima on the line $\vec{\eta}={\bf k}$, i.e. on the
classical phase trajectory of the free moving particle with momentum $k$.

It is worth saying that the Wigner density matrix in the same state with the energy 
$E={\bf k}^2/2$ takes the form
\begin{eqnarray}
\rho_w(\vec{\xi},\vec{\eta})=\delta({\bf k}-\vec{\eta}).
\end{eqnarray}
Therefore, it corresponds to the classical limit of the Fock--Bargmann density matrix.

Let us now turn to the example of the harmonic oscillator which has a discrete spectrum only.
In order to simplify the situation while keeping valid all its main features, we consider the 
one-dimensional case first. Then $R$ and $S$  are scalar complex variables,
$$R={1\over\sqrt{2}}(\xi+i\eta),~~S=R^*={1\over\sqrt{2}}(\xi-i\eta),$$
where $\xi$ and $\eta$ are the coordinate and momentum, respectively. The general 
one-dimensional density matrix $\exp(RS)$, 
similarly to (\ref{a5}), can be expanded over the states with either given coordinate $x$, or
given momentum $k$, or over the one-dimensional harmonic-oscillator states. We are interested in the latter 
expansion,
\begin{eqnarray}
\label{a8}
\exp(RS)=\sum_{n=0}^\infty~{1\over n!}(RS)^n.
\end{eqnarray}
Eq. (\ref{a8}) is followed by an expression for the orthonormalized (with the Bargmann measure) 
wave functions $\phi_n(R)$ of the one-dimensional harmonic oscillator\cite{Fil3},
\begin{eqnarray}
\label{b3}
\phi_n(R)={1\over\sqrt{n!}}~R^n,
\end{eqnarray}
where $n$ is the number of excitation quanta. As for the harmonic-oscillator Hamiltonian 
$\hat{H}_{\mbox{osc}}$ in the Fock--Bargmann space, it is 
\begin{eqnarray}
\hat{H}_{\mbox{osc}}=R{d\over dR}+{1\over2}.
\end{eqnarray}
Due to Eq. (\ref{a2}), the coordinate operator is
\begin{eqnarray}
\label{a9}
\hat{x}={1\over\sqrt{2}}(R+{d\over dR}),
\end{eqnarray}
and due to (\ref{c1}), the one-dimensional momentum is
\begin{eqnarray}
\label{a10}
\hat{k}= -{i\over\sqrt{2}}(R-{d\over dR}),
\end{eqnarray}
and the kinetic energy operator is
\begin{eqnarray}
\label{a11}
\hat{T}= {1\over2}\hat{k}^2=-{1\over4}(R-{d\over dR})^2.
\end{eqnarray}
Evidently,
\begin{eqnarray*}
\hat{H}_{\mbox{osc}}=\hat{T}+{1\over2}~\hat{x}^2.
\end{eqnarray*}

The probability distribution for the harmonic-oscillator with the number of quanta $n$ takes the form
\begin{eqnarray}
\label{a12}
\rho_n(\xi,\eta)\equiv \phi_n^*(R)\phi_n(R)\exp(-RS) \nonumber \\
={1\over n!}\left({\xi^2+\eta^2\over2}\right)^n\exp\{-{\xi^2+\eta^2\over2}\}.
\end{eqnarray}
It takes only positive values and is concentrated around the circle
\begin{eqnarray}
\label{z}
{\xi^2+\eta^2\over2}=n,
\end{eqnarray}
which is the phase trajectory of the classical oscillator with the energy $n$. Evidently,
the left-hand-side part of Eq.(\ref{z}) is the classical Hamilton function of the one-dimensional oscillator.

Let us consider in more detail the limiting case of large $n$. At $n\gg1$, the following 
asymptotic formula holds,
\begin{eqnarray}
\label{z1}
\rho_n(\xi,\eta)\sim{1\over\sqrt{2\pi n}}
\exp\left\{-{1\over2n}\left({\xi^2+\eta^2\over2}-n\right)^2\right\}.
\end{eqnarray}
This is the quantum distribution function for the one-dimensional 
harmonic-oscillator at 
$n\gg1$. It makes us possible to calculate the average value of the classical 
Hamilton function and its dispersion,
$$\overline{\xi^2+\eta^2\over2}=n,~~
\sqrt{\overline{\left({\xi^2+\eta^2\over2}-n\right)^2}}=\sqrt{n}.$$

In order to underline the problems with the Wigner function $\rho_w(\xi,\eta)$,
we define it following the usual way for the harmonic-oscillator with $n=1$. Then,
\begin{eqnarray}
\label{a13}
\rho_w(\xi,\eta)={\xi^2+\eta^2-1\over2}\exp\{-{\xi^2+\eta^2\over2}\}.
\end{eqnarray}
The latter distribution takes negative values inside the circle 
$\xi^2+\eta^2=1$, which, of course, is a drawback. At large values of $\xi^2+\eta^2$, though, 
the distributions $\rho_1(\xi,\eta)$ and $\rho_w(\xi,\eta)$ are identical.

\subsection{Three-dimensional oscillator}

In the three-dimesnional harmonic-oscillator case, the distribution (\ref{a8}) is somewhat modified,
\begin{eqnarray}
\label{x1}
\exp({\bf RS})=\sum_{n=0}^\infty~{1\over n!}({\bf RS})^n.
\end{eqnarray}
Again, $n$ is the number of excitation quanta. However, now $({\bf RS})^n$ are wave packets 
of basis states with the indices of SU(3) symmetry $(n,0)$. In order to unambiguiously identify 
these basis states, we introduce two more quantum numbers, the angular momentum $l$ and its projection $m$. Then,
\begin{eqnarray}
\label{x2}
\exp({\bf RS})=\sum_{n=0}^\infty\sum_{l,m}~W_{n,l}
{1\over n!}(RS)^n D^l_{0,m}(\Omega) D^l_{m,0}(\Omega^*),\nonumber \\
\end{eqnarray}
where the diagonal elements of the density matrix 
\begin{eqnarray}
W_{n,l}={n!(2l+1)\over(n-l)!!(n+l+1)!!},
\end{eqnarray}
and $D_{J}^{M,K}$ are Wigner $D$-functions.
Besides,
$$ R^2= R_x^2+ R_y^2+ R_z^2;~~
{\bf R}=( R_x, R_y, R_z).$$
Therefore, the product $R^n D^l_{0,m}(\Omega)$ 
is the homogeneous harmonic polynomial of the $n$th degree.

We note that the following condition is satisfied,
$$W_{n,l}=1,$$
as long as the parities of $l$ and $n$ are the same and $l\le n$.

In the limiting case of $n\gg l\gg 1$,
\begin{eqnarray}
\label{x3}
W_{n,l}\sim
\rightarrow{2l+1\over n}\exp\{-{l(l+1)\over2n}\}.
\end{eqnarray}

We have arrived to the density matrix for the states with $n$ quanta and the angular momentum 
$l$. It has the form similar to the Gibbs distribution for the rotator with the momentum of inertia $n$ and $kT=\hbar\omega$, where $\omega$ is the harmonic-oscillator frequency.

\subsection{Two three-dimensional oscillators}

Two three-dimensional oscillators produce a manyfold of such diagonal matrix elements of the 
density matrix, which is better classified by the reduction SU(3) $\times$ SU(3) $\to$ SU(3).
$$
\exp\{({\bf R}_1{\bf S}_1)+({\bf R}_2{\bf S}_2)\}   
$$
\begin{eqnarray}
\label{x4}
=\sum_{n_1=0}^\infty\sum_{n_2=0}^\infty~{1\over {n_1}!{n_2}!}
({\bf R}_1{\bf S}_1)^{n_1}({\bf R}_2{\bf S}_2)^{n_2}. 
\end{eqnarray}
Then, the product of wave packets with SU(3) symmetry $(n_1, 0)$ and $(n_2, 0)$ is convenient
to present as a superposition of normalized wave packets $F^{(n_1+n_2-2\mu,\mu)}_{(n_1,0)(n_2,0)}$, with the SU(3) symmetry $(n_1+n_2-2\mu,\mu)$.
$$
{1\over {n_1}!{n_2}!}
({\bf R}_1{\bf S}_1)^{n_1}({\bf R}_2{\bf S}_2)^{n_2}
$$
\begin{eqnarray}
=\sum_\mu W^{(n_1+n_2-2\mu,\mu)}_{(n_1,0)(n_2,0)}
F^{(n_1+n_2-2\mu,\mu)}_{(n_1,0)(n_2,0)},
\end{eqnarray}
$$
W^{(n_1+n_2-2\mu,\mu)}_{(n_1,0)(n_2,0)}=N(n_1,n_2)
$$
\begin{eqnarray}
\times
{n_1!n_2!(n_1+n_2-2\mu+1)!\over\mu!(n-\mu)!(n_2-\mu)!(n_1+n_2-\mu+1)!},
\end{eqnarray}
where $N(n_1n_2)$ is determined by the normalization condition,
\begin{eqnarray}
\sum_{\mu=0}^{\min(n_1,n_2)}W^{(n_1+n_2-2\mu,\mu)}_{(n_1,0)(n_2,0)}=1.
\end{eqnarray}
If $n_1,n_2\gg\mu\gg1$, the expression for the diagonal matrix elements of the density matrix 
is essentially simplified,
\begin{eqnarray}
\label{x5}
W^{(n_1+n_2-2\mu,\mu)}_{(n_1,0)(n_2,0)}\sim {1\over\mu!}
\left({n_1n_2\over{n_1+n_2}}\right)^\mu
\exp\left(-{n_1n_2\over{n_1+n_2}}\right).\nonumber \\
\end{eqnarray}

Again, like in the previous subsection, we have arrived to a conventional in statistical physics 
form of the distribution function, but this time, for the states with various SU(3) symmetry. It 
is equally important that the density matrix elements are squared Clebsch--Gordan coefficients 
of the SU(3) group, whilst the limiting relation (\ref{x5}) provides the asymptotic form of these coefficients.

\subsection{Antisymmetrization effects}

Keeping the story within the limits of pictorial one-dimensional systems, let us turn to the influence of the antisymmetrization to the probability distribution. The simplest case would be two identical particles. With the use of two orbitals (\ref{a1}), we perform the antisymmetrization and separate out the center-of-mass motion, thus getting the following 
orbital in the c.o.m. frame,
\begin{eqnarray}
\phi_{x}^-(R)={1\over\sqrt{2}}\left(\phi_{x}(R)-\phi_{x}(-R)\right),
\end{eqnarray}
$$x={x_1-x_2\over\sqrt{2}},~~R={R_1-R_2\over\sqrt{2}}.$$
$\phi_{x}^-(R)$ 
is the eigenfunction of the operator
\begin{eqnarray}
\hat{x^2}={1\over2}(R+{d\over dR})^2,
\end{eqnarray}
with the eigenvalue $x^2$. The probability distribution for this antisymmetric wave function in 
the Fock--Bargmann space is 
\begin{eqnarray}
{\phi_{x}^-}^*(R)\phi_{x}^-(R)\exp(-RS) 
\end{eqnarray}
$$
=\pi^{-1/2}\left(\cosh(2x\xi)-\cos(2x\eta)\right)\exp(-x^2-\xi^2).
$$

\begin{figure}
\includegraphics[angle=90,width=8.5cm]{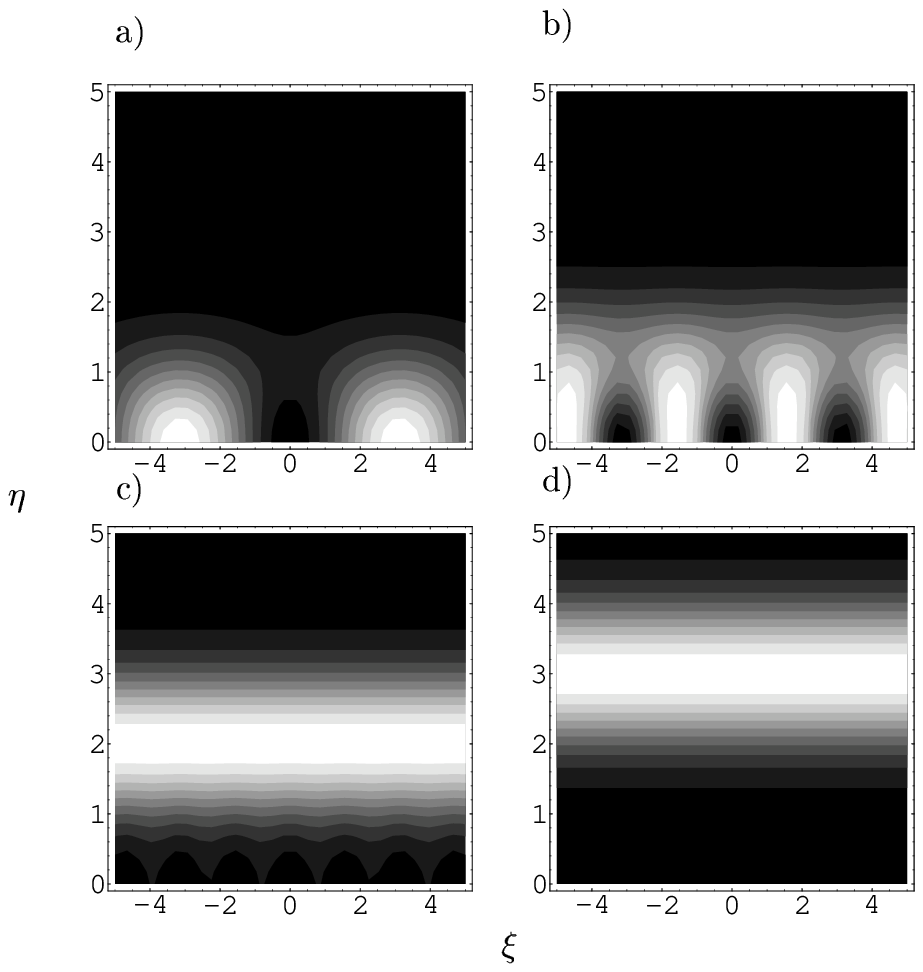}
\caption{Probability distribution $\rho_{k}^-(\xi,\eta)$ for the singlet two-nucleon system, at (a) $k=0.5$, (b) $k=1.0$,
(c) $k=2.0$, and (d) $k=3.0$. A brighter area corresponds to a highter probability.}
\vspace{2cm}
\includegraphics[angle=90,width=8.5cm]{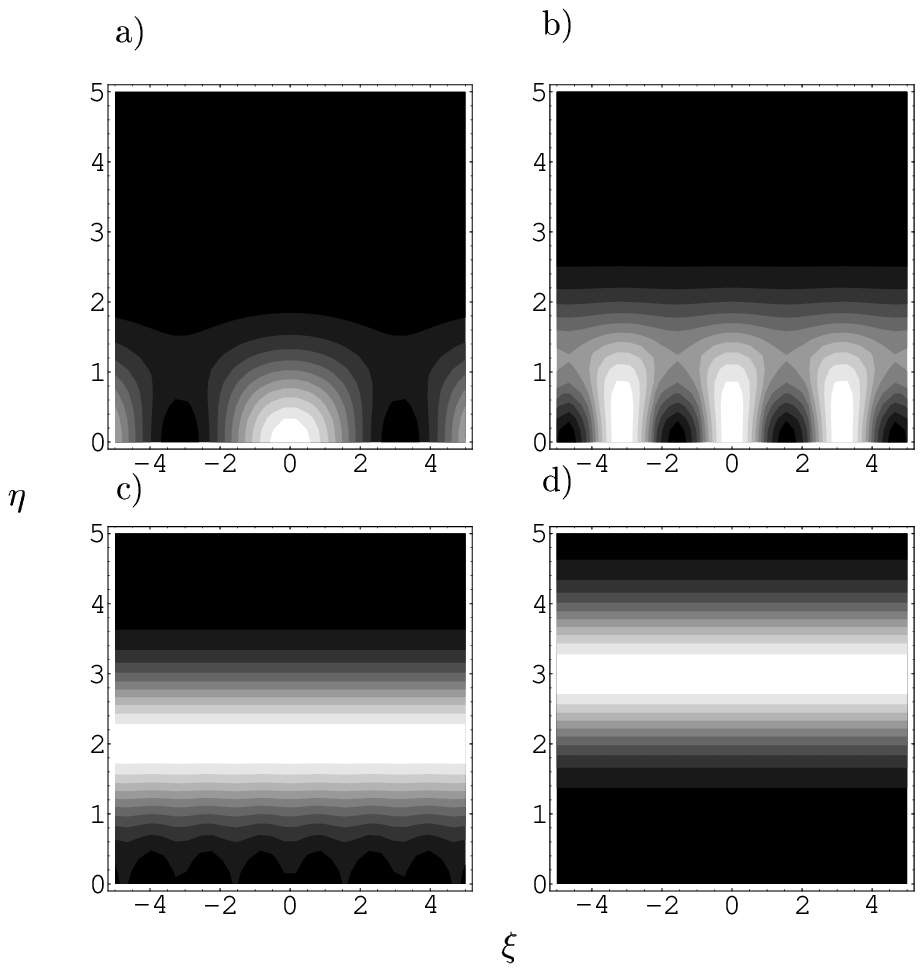}
\caption{Probability distribution $\rho_{k}^+(\xi,\eta)$ for the triplet two-nucleon system, at (a) $k=0.5$, (b) $k=1.0$,
(c) $k=2.0$, and (d) $k=3.0$.}

\end{figure}

Similar transformations occur due to the antisymmetrization in the plane wave case,
\begin{eqnarray}
\phi_{k}^-(R)={1\over\sqrt{2}}\left(\phi_{k}(R)-\phi_{k}(-R)\right),
\end{eqnarray}
as well as in the free motion case,
\begin{eqnarray}
\rho_{k}^-(\xi,\eta)={\phi_{k}^-}^*(R)\phi_{k}^-(R)\exp(-RS)
\end{eqnarray}
\begin{eqnarray*}
\label{y1}
=\pi^{-1/2}\left(\cosh(2k\eta)-\cos(2k\xi)\right)\exp(-k^2-\eta^2).
\end{eqnarray*}

Only at $k\gg 1$ does the latter distribution limit to a superposition of the 
previously found distribution $\rho_{k}(R)$ and $\rho_{-k}(R)$. At low values 
of $k$ an $\eta$-dependence appear. The surfaces $\rho_{k}^-(\xi,\eta)$ for 
several values of $k$ are shown in Fig.~1.

The symmetrization of the wave function of two identical particles leads to the 
following expression for the probability distribution,
\begin{eqnarray}
\label{y2}
\rho_{k}^+(\xi,\eta)
=\pi^{-1/2}\left(\cosh(2k\eta)+\cos(2k\xi)\right)\exp(-k^2-\eta^2). \nonumber \\
\end{eqnarray}
The surface described by Eq.~(\ref{y2}) is shown in Fig.~2.

Thus, both the symmetrization and the antisymmetrization significantly affects the behavior of the distribution functions in the vicinity of the origin at low values of the free motion 
energy $k^2/2$. In particular, if $\eta=0$, and $|\xi|$ increases, both functions oscillate 
around the value  $2\pi^{-1/2}\exp(-k^2)$ with the period $2\pi/k$, remaining positive. 

The density matrix of the antisymmetric states multiplied by the Bargmann measure is the 
result of the integration of  $\rho_{k}^-(\xi,\eta)$ over $k$, or $\rho_{x}^-(\xi,\eta)$ over $x$,
$$
w^+(\xi,\eta)\equiv\int\rho_{x}^-(\xi,\eta)dx=\sinh(RS)\exp(-RS)
$$
\begin{eqnarray}
={1-\exp(-2RS)\over2}.
\end{eqnarray}
The antisymmetrization suppresses the probability distribution in the vicinity of the origin, but in the outer region the probability limits to $1/2$ (Fig.~3(a)). 

\begin{figure}
\begin{center}
\includegraphics{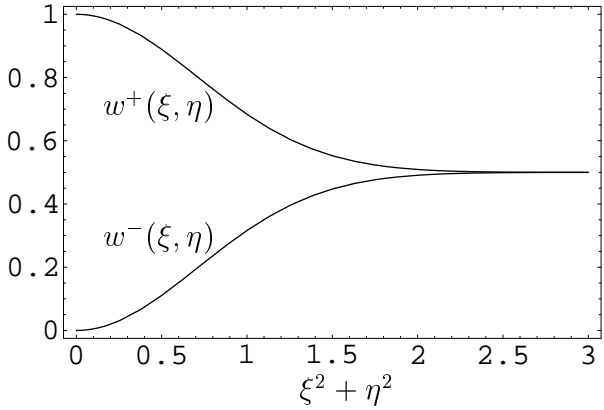}
\end{center}
\caption{Density matrices for the singlet (a) and triplet (b) two-nucleon systems (see Eqs.(2.41--2.42) for definitions).}
\end{figure}

The integration of $\rho_{x}^+(\xi,\eta)$  yields the density matrix of the symmetric states multiplied by the Bargmann measure,
$$
w^+(\xi,\eta)\equiv\int\rho_{x}^+(\xi,\eta)dx=\cosh(RS)\exp(-RS)
$$
\begin{eqnarray}
={1+\exp(-2RS)\over2},
\end{eqnarray}
which favours small values of $\xi,\eta$, and has the same asymptotic limit of $1/2$ (Fig.~4(b)).

The probability distribution for the antisymmetric harmonic-oscillator states with the number of excitation 
quanta $2n+1$ is
\begin{eqnarray}
\rho_{2n+1}^-(\xi,\eta)=
{1\over(2n+1)!}\left({\xi^2+\eta^2\over2}\right)^{2n+1}
\exp\{-{\xi^2+\eta^2\over2}\}.\nonumber \\
\end{eqnarray}
This function reaches its maxima incide a ring, which at large $n$ shrinks to the circle
$${\xi^2+\eta^2\over2}=2n+1,$$
which is the classical trajectory at the energy $2n+1$.

\subsection{Three one-dimensional fermions}

Consider now a system of three one-dimensional fermions with parallel spins. Let 
$x_1,x_2,x_3$ be their coordinates. Besides, let 
$$y_1={1\over\sqrt{2}}(x_1-x_2),~
y_2=\sqrt{{2\over3}}\left(x_3-{x_1+x_2\over2}\right).$$
We replace the three independent Fock--Bargmann variables $R_1,R_2,R_3$ by 
\begin{eqnarray}
\label{f1}
P=\sqrt{{1\over2}}(R_1-R_2)=A\cos\gamma=\xi_1+i\eta_1,
\end{eqnarray}
\begin{eqnarray}
Q=\sqrt{{2\over3}}\left(R_3-{R_1+R_2\over2}\right)=A\sin\gamma=\xi_2+i\eta_2,\nonumber \\
\end{eqnarray}
$$P^*=\sqrt{{1\over2}}(S_1-S_2),~
Q^*=\sqrt{{2\over3}}\left(S_3-{S_1+S_2\over2}\right).$$
Then, the overlap integral of two Slater determinants, 
$$\phi_{y_1,y_2}(P,Q)~\mbox{and}~\phi^*_{y_1,y_2}(P,Q),$$
constructed of Brink--Bloch orbitals in the c.o.m. system, takes the form
\begin{eqnarray*}
\int dy_1\int dy_2\phi^*_{y_1,y_2}(P,Q)\phi_{y_1,y_2}(P,Q)=
\end{eqnarray*}
%\begin{eqnarray*}
%={1\over6}\{\exp\{PP^*+QQ^*\}+
%\exp\{(-{1\over2}P-{\sqrt{3}\over2}Q)P^*+
%({\sqrt{3}\over2}P-{1\over2}Q)Q^*\}+
%\end{eqnarray*}
%\begin{eqnarray*}
%+\exp\{(-{1\over2}P+{\sqrt{3}\over2}Q)P^*+({\sqrt{3}\over2}P-{1\over2}Q)Q^*\}
%-\exp\{-PP^*+QQ^*\}-
%\end{eqnarray*}
%\begin{eqnarray*}
%-\exp\{({1\over2}P-{\sqrt{3}\over2}Q)P^*+
%({\sqrt{3}\over2}P-{1\over2}Q)Q^*\}-
%\end{eqnarray*}
%\begin{eqnarray*}
%-\exp\{({1\over2}P+{\sqrt{3}\over2}Q)P^*+({\sqrt{3}\over2}P-{1\over2}Q)Q^*\}\}=
%\end{eqnarray*}
\begin{eqnarray*}
=\sum_n \sum_{m=0}^{[(n-3)/3]}
\frac{(AA^*)^n
2\cdot\cos(6m+3)\gamma\cos(6m+3)\gamma^*}{(n-6m-3)!!(n+6m+3)!!}+
\end{eqnarray*}
\begin{eqnarray}
\label{e1}
+\sum_n\sum_{m=1}^{[n/6]}
{(AA^*)^n 2\cdot\sin6m\gamma\sin6m\gamma^*\over(n-6m)!!(n+6m)!!}.
\end{eqnarray}

%\end{widetext}

The expansion (\ref{e1}) yields the diagonal elements of the density matrix in the harmonic-oscillator representation.
$$
w_{n,m}'(\xi,\eta_1,\xi_2,\eta_2)=
$$
\begin{eqnarray}
= 2 {(AA^*)^n \cos(6m+3)\gamma\cos(6m+3)\gamma^*\over(n-6m-3)!!(n+6m+3)!!}
\end{eqnarray}
\begin{eqnarray}
w_{n,m}''(\xi,\eta_1,\xi_2,\eta_2)=
2 {(AA^*)^n \sin6m\gamma\sin6m\gamma^*\over(n-6m)!!(n+6m)!!}. \nonumber \\
\end{eqnarray}

The number of quanta $n$ takes only odd values beginning from $n=3$ due to the Pauli principle.
Due to the same principle, the factors at 
$\gamma~(\gamma^*)$ in the arguments of sines and cosines are proportional to three. Again, 
if $n\gg m\gg1$, we arrive to a simple limiting form,
$$
\int dy_1\int dy_2\phi^*_{y_1,y_2}(P,Q)\phi_{y_1,y_2}(P,Q)\sim
$$
$$
\sum_n \sum_m
{(AA^*)^n\over n!}\cdot{2\over n}\exp\left(-{(6m+3)^2\over 2n}\right)
$$
\begin{equation}
\times \cos(6m+3)\gamma\cos(6m+3)\gamma^*  
\end{equation}
$$
+\sum_n\sum_m
{(AA^*)^n\over n!}\cdot{2\over n}\exp\left(-{(6m)^2\over 2n}\right)
\sin6m\gamma\sin6m\gamma^*
$$
which is similar to the one for the three-dimensional oscillator.

With the density matrix $w^i_{n,m}(\xi_1,\eta_1;\xi_2,\eta_2)$ at our disposal, it is easy to find the single-particle density matrix 
$w^i_{n,m}(\xi_1,\eta_1)$ in the form of the integration,
\begin{eqnarray}
w^i_{n,m}(\xi_1,\eta_1)=\int w^i_{n,m}(\xi_1,\eta_1;\xi_2,\eta_2)
\nonumber \\
\times \exp\{-{\xi^2_2+\eta^2_2\over2}\}{d\xi_2d\eta_2\over2\pi}.
\end{eqnarray}

\subsection{Same system, three-dimensional case}

In the case of three three-dimensional fermions having the same direction of spin, the first 
two terms of the expansion of the overlap integral take the form
\begin{eqnarray*}
{1\over2}([{\bf PQ}][{\bf P}^*{\bf Q}^*])+
\end{eqnarray*}
\begin{eqnarray*}
+{1\over24}\{({\bf PP}^*)^3-3({\bf PP}^*)({\bf QP}^*)^2-
3({\bf PP}^*)({\bf PQ}^*)^2+
\end{eqnarray*}
\begin{eqnarray}
\label{x10}
+9({\bf PP}^*)({\bf QQ}^*)^2-
9([{\bf PQ}][({\bf P}^*{\bf Q}^*])({\bf QQ}^*)\}.
\end{eqnarray}
These terms are related to the overlap integrals of two SU(3) irreducible representations 
with the indices (0,1) and (3,0), respectively. It is worth to note that the (1,1) 
irrep is absent.

The first terms of the expansion of the overlap integral determine its behavior 
at small values of the vectors ${\bf P},{\bf Q};{\bf P}^*,{\bf Q}^*.$
Thus, the very first term, 
$${1\over2}([{\bf PQ}][{\bf P}^*{\bf Q}^*]),$$
is the overlap of the translation-invariant wave functions of 
the Elliott's SU(3) model \cite{Ell} of three fermions with the 
same direction of their spins, while the fourth power of this term 
is the overlap for $^{12}$C, in the same model.
If the vectors ${\bf P}$ and ${\bf Q}$ are collinear and are of the same length,
we can set ${\bf P}={\bf Q}={\bf R}.$
Then, the first term of the expansion (\ref{x10}) vanishes, and the second ttakes the 
form
$${1\over6}({\bf RR}^*)^3,$$
which is responsible for the linear structure of the wave function. Again, this term 
corresponds to the overlap integral of the three spin-correlated fermions but occupying 
the states 
$$[0,0,0][1,0,0][2,0,0]$$
of the $s$-,$p$- and $sd-$shells, respectively. Here we use the traditional notation 
$[n_x,n_y,n_z]$ for the single-particle harmonic-oscillator states with the numbers of 
quanta $n_i$ along the axis $i$.

\section{Phase trajectories}

The orbitals (\ref{a1}) clear the way of determining the classical phase trajectories of the wave packets of a system of $A$
fermions. In practice, in order to derive the equations of classical dynamics and find the phase trajectories, several stages
are needed to be gone through. At first, the Slater determinants are constructed of the orbitals 
(\ref{a1}) thus meeting the Pauli principle reqirements. Then, these determinants are declared 
as the trial functions of the variational problem, with the complex vectors
$${\bf R}_i,~~i=1,2,...\mbox{A}$$
being the variational parameters depending on time $t$. Next stage is the calculation of the 
Lagrange function. Finally, the least action principle is applied, which results in the classical equations for
$${\bf R}(t)={\vec{\xi}(t)+i\vec{\eta}(t)\over\sqrt{2}}$$
and, in the end, the phase trajectories are found.

At a given energy, each trajectory of the system of fermions is located on a hyper-surface in 
the phase space. It is completely defined by the integrals of motion. On the other hand, in 
the Fock--Bargmann space, the squared absolute value of the wave function of the system multiplied by the Bargmann measure gives the probability density distribution in the same phase 
space. By comparing the phase trajectories and the quantum distributions, we may judge, to what 
degree and under what conditions the phase trajectories reproduce the real situation.

\subsection{Motion in one dimension}

It is especially easy to find the classic trajectories of motion in one dimension using Eq.~
(\ref{a1}). The following relation is used,
\begin{eqnarray}
\label{a14}
{\cal H}(\xi,\eta)=E,~~
{\langle\phi_{x}^*(R)|\hat{{\mbox{H}}}|\phi_{x}(R)\rangle \over\langle\phi_{x}^*(R)|\phi_{x}(R)\rangle }\equiv
{\cal H}(\xi,\eta).
\end{eqnarray}
Here  $\hat{{\mbox{H}}}$ is the quantum Hamiltonian of a one-dimensional system in the coordinate representation, ${\cal H}(\xi,\eta)$ is the Hamilton function defined as the 
result of variation of the Hamiltonian over the Brink--Bloch orbitals. Thus, for the one-dimensional free motion, when the Hamiltonian is
$$\hat{\mbox{H}}=-{1\over2}{d^2\over dx^2},$$
the Hamilton function appears to be
\begin{eqnarray}
\label{a15}
{\cal H}(\xi,\eta)={1\over2}\eta^2+{1\over4}.
\end{eqnarray}
It follows from (\ref{a15}) that the wave packet (\ref{a1}) moves as a free particle having momentum $\eta$ and the mass of the nucleon. However, as 
$E={\cal H}(\xi,\eta)$, the total energy of the wave packet contains an additional term 
equal to $-1/4$. It is this energy which is required to create a one-dimensional wave packet.
As any wave packet, this one cannot exist for a long time and should spread apart. As far as we neglect this phenomenon and do not consider its mechanism, the minimal energy of the wave 
packet appears to be $1/4$. It remains so even when an attractive potential is switched on 
and a spectrum of states with positive energy not more than $1/4$ appears. Naturally, the quantum description does not have such a drawback.
 
As for the phase trajectories of free motion (\ref{a15}), the dispersion relation 
$$\eta=\sqrt{2E-1/2},$$
somewhat differs from both the classical one and the one following the line of maxima of the distribution function (\ref{b2}).

For the one-dimensional oscillator
\begin{eqnarray}
\hat{\mbox{H}}=-{1\over2}{d^2\over dx^2}+{1\over2}x^2,
\end{eqnarray}
therefore,

\begin{eqnarray}
\label{a16}
{\cal H}(\xi,\eta)={1\over2}(\eta^2+\xi^2)+{1\over2}
\end{eqnarray}
and the phase trajectories at energy $E$ satisfy the relation

\begin{eqnarray}
{1\over2}(\eta^2+\xi^2)+{1\over2}=E.
\end{eqnarray}
As expected, at any $E$ the motion is finite while the spectrum is continuous but 
limited ($E>1/2$). At $E=1/2$ the phase trajectory degenerates into the point of origin ($\xi=0,~\eta=0$). We remind that the quantum distribution in the same state (see Eq. (\ref{a12})) is 
$$\rho_0(\xi,\eta)=\pi^{-1/2}\exp\{-{\xi^2+\eta^2\over2}\}.$$
It has a maximum at the origin point.

\subsection{Free motion of triplet pair of fermions}

In order to understand how the anstisymmetrization affects the phase trajectories of free one-dimensional motion of two fermions in the triplet state, we define the classical Hamilton function,
\begin{eqnarray}
{\cal H}^-(\xi,\eta)=
{\langle\phi_{x}^{-*}(R)|\hat{{\mbox{H}}}|\phi_{x}^-(R)\rangle \over
\langle\phi_{x}^{-*}(R)|\phi_{x}^-(R)\rangle },
\end{eqnarray}
where, again,
$$\hat{\mbox{H}}=-{1\over2}{d^2\over dx^2}.$$
Simple calculations lead to the following result \cite{Fil1},
\begin{eqnarray}
\label{u}
{\cal H}^-(\xi,\eta)={\eta^2\over2}+{1\over4}+{\xi^2+\eta^2\over4}
\left(\coth{\xi^2+\eta^2\over2}-1\right). \nonumber \\
\end{eqnarray}

\begin{figure}
\begin{center}
%\vspace{-2cm}
\includegraphics{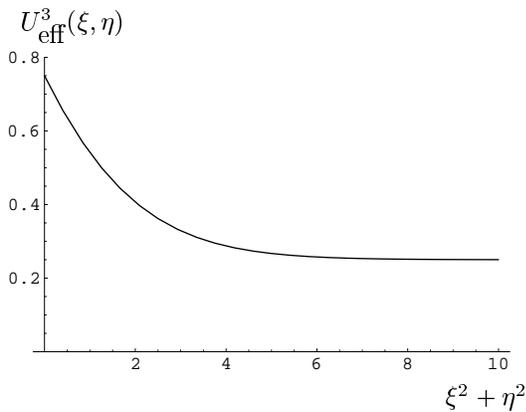}
%\epsfbox{fig4tex.eps}
\end{center}
\caption{Effective potential in the triplet fermion pair case.}
\end{figure}

\begin{figure}
\begin{center}
%\vspace{-2cm}
\includegraphics{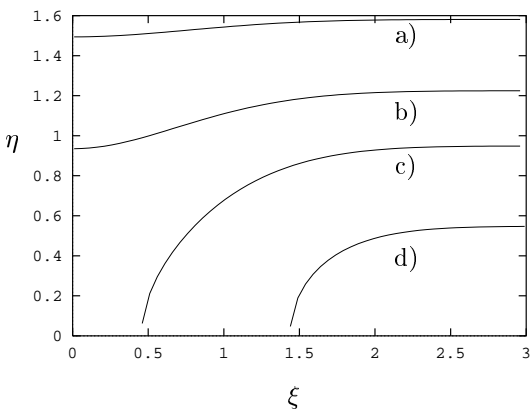}
%\epsfbox{fig5tex.eps}
\end{center}
\caption{Phase trajectories in the triplet fermion pair case, a) $E=1.5$, b) $E=1.0$, c) $E=0.7$, d) $E=0.4$. }
\end{figure}

The sum of two last terms in the right-hand side of (\ref{u})
is shown in Fig.~4. Having set 
${\cal H}^-(\xi,\eta)=E,$ we obtain the phase trajectory at energy $E$. Note that the 
minimal energy is equal to $1/4$. Besides, as
$$\coth{\xi^2+\eta^2\over2}-1\geq 0,$$
the Hamilton function contains a repulsion. Therefore, as long as $1/4\leq E\leq3/4,$ 
the trajectory begins at large positive values of $\xi$ and $\eta=\sqrt{2E-1/2}$
and after the reflection does not enter the region of negative $\xi$. There appears a turning point where the momentum is zero. Having passed through this point, the trajectory remains in the region of positive $\xi$ with the sign of momentum alternated (Fig.~5). If 
$E\geq3/4,$ the motion never stops but slows down. Then the trajectory enters the region of negative $\xi$ where it can be reproduced by a reflection around the $\eta$ axis.

The repulsion visible in the phase trajectories of the triplet pair of fermions at small values of $\xi$ and $\eta$ agrees well with the conventional wisdom: the Pauli principle forbids the fermions having parallel spins to be located at the same point.

The probability distribution (\ref{y1}) for the triplet pair with energy $E=k^2/2$ limits to 
the phase trajectory 
${\cal H}^-(\xi,\eta)=E$ only if 
$E\gg3/4,$ i.e., when the influence of the Pauli principle and the antisymmetrization is negligible. It is then when the classical result is the asymptotic limit of the quantum one.
However, at small energies, where the corrections to the classical results due to the antisymmetrization become significant, there is no analogy between the quantum probability 
distribution and the phase trajectories. In order for this analogy to appear, some additional corrections of classical equations are required. These corrections are mostly due to the spreading of wave packets described by classical equations.

\subsection{Singlet pair}

In the case of the singlet pair of free fermions, the Hamilton function takes the form
\begin{eqnarray}
\label{u1}
{\cal H}^+(\xi,\eta)={\eta^2\over2}+{1\over4}+{\xi^2+\eta^2\over4}
\left(\tanh{\xi^2+\eta^2\over2}-1\right). \nonumber \\
\end{eqnarray}
The last term in the r.h.s. of (\ref{u1}) appears, again, due to the symmetrization but, 
unlike in (\ref{u}), it corresponds to the attraction in the vicinity of the origin of the 
phase space, as
$$\tanh{\xi^2+\eta^2\over2}-1\leq0.$$

\begin{figure}
\begin{center}
\includegraphics{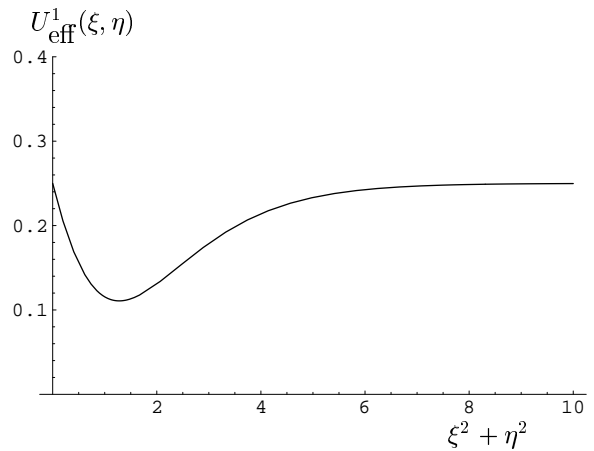}
%\epsfbox{fig6tex.eps}
\end{center}
\caption{Effective potential in the singlet fermion pair case.}
\end{figure}

The behavior of two last terms in the right-hand side of (\ref{u1}) is shown in Fig.~6. The minimum
of this sum is located at the circle centered around the origin. Having added $\eta^2/2$ to 
this sum, we get two minima of the Hamilton function at
$$\xi=\pm \xi_0,~~\xi_0=1.13,~~\eta=0,$$
when $E_{\min}=0.11.$
In these minima, the phase trajectories degenerate into a point. Then, when the energy exceeds 
its minimal value, there appear closed phase trajectories of finite motion (Fig.~7). The motion becomes 
infinite if $E>1/4$. However, an explicit relation between the infinite trajectories and the quantum probability distribution (\ref{y2}) becomes evident only if 
$E\gg1/4$ and the influence of the symmetrization in both the phase trajectories and 
probability distributions becomes negligible. As for the finite trajectories, they are the 
consequence of the approximation applied at the derivation of the classical equations. The approximation under question is that the wave packets are considered as having a fixed width,
although in fact they spread apart as they move. This phenomenon may be neglected at high
energies only.

\begin{figure}
\begin{center}
\includegraphics{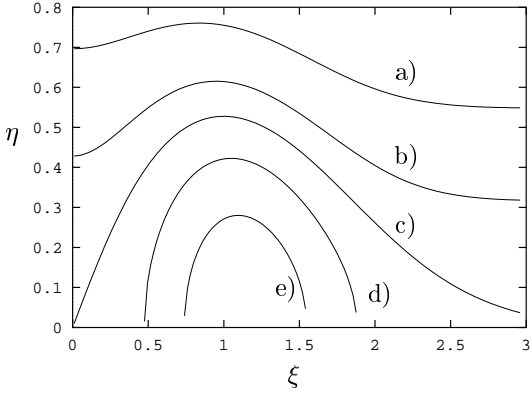}
%\epsfbox{fig7tex.eps}
\end{center}
\caption{Phase trajectories in the singlet fermion pair case, a) $E=0.40$, b) $E=0.30$, c) $E=0.25$, d) $E=0.25$, and e) $E=0.15$.}
\end{figure}

\section{Interacting fermions}

We now introduce a fermion-fermion interaction potential in order to study those problems which 
appear when a system has both discrete and continuous spectra. The interaction will be simulated 
by a Gaussian interactive potential,

\begin{eqnarray}
\label{g1}
U(x)=-V_0\exp\{-\alpha x^2\}.
\end{eqnarray}
First, we find the spectrum of states defined in the Fock--Bargmann space, then we perform a transformation to the classical equations and determine the phase trajectories.

\subsection{Solution of the quantum wave equation}

In the framework of quantum approach, we find the overlap integral

$$\langle S|U(x)|R\rangle =\int \phi_{x}( S)U(x)\phi_{x}( R)dx$$
of the potential energy operator (\ref{g1}) between the one-dimensional Brink--Bloch orbitals
\begin{eqnarray}
\label{h1}
\phi_{x}( R)=\pi^{-1/4}\exp\{-{x^2\over2}+\sqrt{2}Rx-{R^2\over2}\},
\end{eqnarray}
\begin{eqnarray}
\label{h2}
\phi_{x}( S)=\pi^{-1/4}\exp\{-{x^2\over2}+\sqrt{2}Sx-{S^2\over2}\}.
\end{eqnarray}
It is easy to see that
\begin{eqnarray}
\label{h3}
\langle S|U(x)|R\rangle =-z^{1/2}V_0\exp\{zRS+{z-1\over2}(R^2+S^2)\}, \nonumber \\
\end{eqnarray}
where $z^{-1}=1+\alpha.$
A similar overlap with the kinetic energy operator has been calculated earlier,
\begin{eqnarray}
\label{h4}
\langle S|\hat{T}|R\rangle =\left(-{1\over4}(R-S)^2+{1\over4}\right)\exp(RS).
\end{eqnarray}

The overlap integrals (\ref{h3}) and (\ref{h4}) are the matrix elements of the operators involved between the orbitals (\ref{h1}) and (\ref{h2}). But these orbitals are superpositions
of all orthonormalized basis functions of one-dimensional harmonic oscillator defined in the 
conventional coordinate space. The coefficients of these superpositions are the basis 
functions of the same oscillator but in the Fock--Bargmann space. For example, for (\ref{h1}),
\begin{eqnarray}
\phi_{x}( R)=\sum_{n=0}^\infty {1\over\sqrt{2^n n!\sqrt{\pi}}}
H_n(x)\exp\{-x^2/2\}{1\over\sqrt{n!}}R^n, \nonumber \\
\end{eqnarray}
where $H_n(x)$ are Hermite polynomials. Having integrated over $x$, we found a 
linear combination of all matrix elements. It is our task now to extract the proper matrix elements 
from this combination. 

A matrix element $\langle S|\hat{O}|R\rangle $ of an operator $\hat{O}$ can be expanded as follows,
\begin{eqnarray}
\label{h5}
\langle S|\hat{O}|R\rangle =\sum_{n=0}^\infty\sum_{\tilde{n}=0}^\infty
{1\over\sqrt{n!}}R^n
\langle n|\hat{O}|\tilde{n}\rangle {1\over\sqrt{\tilde{n}!}}S^{\tilde{n}},\nonumber \\
\end{eqnarray}
where
$$\langle n|\hat{O}|\tilde{n}\rangle =\int_{\infty}^\infty~dx~
{1\over\sqrt{2^n n!\sqrt{\pi}}}$$
$$ \times 
H_n(x)\exp\{-x^2/2\}
\hat{O}{1\over\sqrt{2^{\tilde{n}}\tilde{n}!\sqrt{\pi}}}
H_{\tilde{n}}(x)\exp\{-x^2/2\},$$
which shows the way to solve our task. Indeed, as
\begin{eqnarray}
\langle S|\hat{U}(x)|R\rangle =\sum_{n=0}^\infty\sum_{\tilde{n}=0}^\infty
{1\over\sqrt{n!}}R^n
\langle n|\hat{U}(x)|\tilde{n}\rangle {1\over\sqrt{\tilde{n}!}}S^{\tilde{n}}, \nonumber \\
\end{eqnarray}
the partial matrix elements are
\begin{eqnarray}
\langle n|\hat{U}(x)|\tilde{n}\rangle =
-V_0\sum_{m=0}^{\min(n,\tilde{n})}{n!\tilde{n}!z^{m+1/2}(z-1)^{n+\tilde{n}-2m}\over m!(n-m)!!
(\tilde{n}-m)!!}. \nonumber \\
\end{eqnarray}

Eqs.(\ref{h4}) and (\ref{h5}) are followed by three well-known expressions for the matrix 
elements of the kinetic energy operator of one-dimensional oscillator,
\begin{eqnarray}
\langle n+2|\hat{T}|n\rangle =-{1\over4}\sqrt{(n+1)(n+2)},
\end{eqnarray}
\begin{eqnarray}
\langle n-2|\hat{T}|n\rangle =-{1\over4}\sqrt{(n-1)n},
\end{eqnarray}
\begin{eqnarray}
\langle n|\hat{T}|n\rangle ={1\over2}n+{1\over4}.
\end{eqnarray}
We used the matrix elements between the Brink--Bloch orbits and basis functions defined in the Fock--Bargmann space. We have specifically considered a simplest case when all calculations are free of background details. However, the algorythm outlined above remains applicable in a more complicated situation if the overlap integrals and the basis functions in the Fock--Bargmann representation are known.

Now, using the matrix elements of the Hamiltonian found above, we consider a set of algebraic equations
\begin{eqnarray}
\label{h6}
\sum_{\tilde{n}}^\infty(\langle n|\hat{H}|\tilde{n}\rangle -E\delta_{n,\tilde{n}})C_n=0,~~
n=0,1,2,..., \nonumber \\
\end{eqnarray}
for the coefficients $C_n$ of the expansion of the wave function $\Psi(R)$ over the harmonic-oscillator basis. Naturally,
\begin{eqnarray}
\label{h7}
\Psi(R)=\sum_{n=0}^\infty C_n{1\over\sqrt{n!}}R^n.
\end{eqnarray}
The same coefficients $C_n$ enter the expansion
\begin{eqnarray}
\label{h8}
\Psi(x)=\sum_{n=0}^\infty C_n{1\over\sqrt{2^n n!\sqrt{\pi}}}
H_n(x)\exp\{-x^2/2\}
\end{eqnarray}
of the wave function in the coordinate representation which has (\ref{h7}) as its image in the Fock--Bargmann space.

The solution of the set (\ref{h6}) yields the spectrum of the potential  $U(x).$ Sometimes the 
applicability of the expansion  (\ref{h7}) to the continuum states is questioned. This is related to the
fact that, in the coordinate representation, one-dimensional continuum wave functions oscillate with $x$ 
increases with a non-decreasing amplitude. Meanwhile, the harmonic-oscillator basis functions fall sharply with the increase of $x$, and the question appears, whether it is possible to expand a non-decreasing oscillating function in a series over the harmonic-oscillator functions. Precisely speaking, the convergence of such a series is under question.

In the defence of the methode, we make a note on the expansion (\ref{h7}), which is a Tailor series of an entire function  $\Psi(R).$ By definition of a whole function, its expansion over powers of $R$ coverges everywhere in the complex plane except the infinity where this function has an irregularity. This general 
statement on the convergence of (\ref{h7}) is confirmed by the behavior of the coefficients of the series: 
$|C_n|$ are limited and $1/\sqrt{n!}$ converges to zero. In a circle of any radius centerd around 
$R=0$ this convergence is uniform. It ceases to be steady out of this circle.

We can analyze  an example of free motion, when 
$U(x)=0$. As in the general case, the continuum is degenerate in the two-fold way, and in order to remove this degeneracy, parity is introduced as an integral of motion along with energy $E$. Evidently, the wave function of an even state
\begin{eqnarray}
\Psi^+(R)=\sum_{n=0}^\infty C^+_n{1\over\sqrt{(2n)!}}R^{2n}
\end{eqnarray}
contains only the basis states proportional to even powers of 
$R,$ and the wave function of an odd state
\begin{eqnarray}
\Psi^-(R)=\sum_{n=0}^\infty C^-_n{1\over\sqrt{(2n+1)!}}R^{2n+1}
\end{eqnarray}
contains those proportional to odd powers of $R$ only. 

The values of the coefficients  $C^+_n$ ($C^-_n$) of the expansion of the wave function of free motion follow the expression
\begin{eqnarray}
\label{h9}
\phi_k(R)=\pi^{-1/4}\exp\{-{k^2\over2}-i\sqrt{2}Rk+{R^2\over2}\}
\end{eqnarray}
which is a one-dimensional analogue of the plane wave (\ref{a6}) with the momentum $k.$
Note that (\ref{h9}) is not only the wave function of free motion in one dimension but also the generating 
function of the Hermite polynomials $H_n(k)$, with $R$ being the generating parameter. Therefore \cite{Bat},
\begin{eqnarray}
\label{h10}
\phi_k(R)=\sum_{n=0}^\infty{i^n\over\sqrt{2^n n!\sqrt{\pi}}}H_n(k)
\exp\{-{k^2\over2}\}{1\over\sqrt{n!}}R^n,\nonumber \\
\end{eqnarray}
\begin{eqnarray}
\label{h11}
C^+_n(k)={(-1)^n\over\sqrt{2^{2n}(2n)!\sqrt{\pi}}}H_{2n}(k)
\exp\{-{k^2\over2}\},\nonumber \\
\end{eqnarray}
\begin{eqnarray}
\label{h12}
C^-_n(k)={(-1)^ni\over\sqrt{2^{2n+1}(2n+1)!\sqrt{\pi}}} H_{2n+1}(k).
\exp\{-{k^2\over2}\}.\nonumber \\
\end{eqnarray}

We underline some important facts here. First, as expected, the coefficients $C^+_n$ and $C^-_n$
satisfy the set  (\ref{h6}) at $U(x)=0$. This is a consequence of the known\cite{Bat} recurrent 
relations for the Hermite polynomials which are just identical to the set (\ref{h6}) in this particular case.
Second, these coefficients are the eigenfunctions of the harmonic oscillator defined in the momentum 
representation. 

Finally, the asymptotic behavior\cite{Bat} of $C^{+}_{n}$ (and $C^{-}_{n}$) in the limiting case 
$n \gg 1$
\begin{eqnarray}
C^+_n(k) \sim (-1)^n {1\over\sqrt{\pi}} n^{-1/4} \cos(k\sqrt{4n+1}),
\end{eqnarray}
\begin{eqnarray}
C^-_n(k)\sim (-1)^n{i\over\sqrt{\pi}}n^{-1/4}\sin(k\sqrt{4n+3})
\end{eqnarray}
is identical to that of the solution of the one-dimensional Schr\"odinger equation in the 
coordinate representation for free motion with energy $E=k^2/2$. Here, the role of the absolute value 
of the coordinate is played by the value of the turning point of motion in the harmonic-oscillator field, i.e. $\sqrt{4n+1},$ 
if energy is 
$2n+1/2,$ and $\sqrt{4n+1},$ when it equals $2n+3/2.$
The expansion coefficients are normalized,
\begin{eqnarray}
\sum_n^\infty C^\pm_n(k)C^\pm_n(k')=\delta(E-E').
\end{eqnarray}

Naturally, the potential  $U(x)$ affects the values of the expansion coefficients but we still can define 
{\it a priori} its asymptotic behavior which takes a familiar form,
\begin{eqnarray}
C^+_n(k)\sim
(-1)^n{1\over\sqrt{\pi}}n^{-1/4}\cos(k\sqrt{4n+1}+\delta^+(k)),\nonumber \\
\end{eqnarray}
\begin{eqnarray}
C^-_n(k)\sim (-1)^n{i\over\sqrt{\pi}}n^{-1/4}\sin(k\sqrt{4n+3}+\delta^-(k))\nonumber \\
\end{eqnarray}
Thus, at any $n\geq n_0$, when the asymptotic expressions are defined with a pre-defined precision,
all coefficients are defined by one unknown parameter only, the phase shift $\delta^+(k)~(\delta^-(k))$, which has to be found along with the coefficients $C_n^+~(C_n^-)$, at $n<n_0.$ In this way, the
infinite set of equations is reduced to a closed set of 
$n_0+1$ equations. The choise of  $n_0$ depends on the desired precision of calculations.

These general considerations basic speculations remain valid in the three-dimensional case, as well as in the
cases of many particle systems, and with the Coulomb interaction included. In any case, it is necessary to
detemine the explicit form of the asymptotic behavior, which, in general, contains the scattering $K$- or $S$-matrix elements, and express it in terms of quantum numbers of the muti-dimensional harmonic 
oscillator\cite{Fil2}.

\subsection{Discussion on solutions}

Let the attractive potential (\ref{g1}) generates a single bound state with energy $-\epsilon$, 
positive parity and the expansion coefficients $\{C_n^\epsilon\}.$ As for the continuum states, 
they are labeled by the momentum at infinity $k$ and parity. The expansion coefficients are $\{C_n^+(k)\}$ for even (singlet) and $\{C_n^{-}(k)\}$ for odd (triplet) states.

Energy and structure of the wave function of the ground state depend on the depth $V_0$ of the potential. If 
$V_0\gg1$ then
\begin{eqnarray}
-\epsilon\sim -V_0+\sqrt{{V_0\over2}},
\end{eqnarray}
and the wave function is localized around the origin in both coordinate and phase representations. If, however, $V_0$ is approaching zero, then $-\epsilon\rightarrow 0,$ and the wave function becomes slowly falling and delocalized, spreading far away from the origin.

The density probability distribution $\rho_\epsilon(\xi,\eta)$ in the phase space %(Fig.~8) 
is defined in terms of a two-fold summation,
$$\rho_\epsilon(\xi,\eta)=\sum_{n=0}^\infty\sum_{\tilde{n}=0}^\infty
C^\epsilon_n{1\over\sqrt{(2n)!}}R^{2n}$$
\begin{eqnarray}
\label{g2}
\times C^\epsilon_{\tilde{n}}{1\over\sqrt{(2\tilde{n})!}}S^{2\tilde{n}}
\exp\{-{\xi^2+\eta^2\over2}\}.
\end{eqnarray}
As usual and as opposite to the classical result ((\ref{u}) and (\ref{u1}))
for the phase trajectories, the infinite motion is possible at any $E\ge0$.

\subsection{Phase trajectories for the potential (\ref{g1})}

If motion of a particle in the field of the potential 
$U(x)=-V_0\exp\{-\alpha x^2\}$ is studied, when the Pauli principle does not affect 
the results in any way, the following term should be added to the Hamilton function of free
motion (\ref{a15}), 
\begin{eqnarray}
\label{g3}
{\langle \phi_{x}^*(R)|U(x)|\phi_{x}(R)\rangle \over\langle \phi_{x}^*(R)|\phi_{x}(R)\rangle }=
-z^{1/2}V_0\exp\{{z-1\over2}(R+S)^2\}. \nonumber \\
\end{eqnarray}
Then,
\begin{equation}
{\cal H}(\xi,\eta)={\eta^2\over2}+{1\over4}-z^{1/2}V_0\exp\{-(1-z)\xi^2\}.
\end{equation}
As in the free motion case, infinite trajectories appear only if $E\geq1/4.$ Note that the shape
of the potential remain Gaussian after averaging over the Brink--Bloch orbitals. However, as 
$1\geq z\geq 0,$ the depth of the resultant potential appears to be 
$\sqrt{z}$ times less than that of the original potential while its width increases $\sqrt{z}$ 
times. Therefore, even in the limiting case of high energy when the term $1/4$ in the expression for the Hamilton function can be neglected, the scattering picture should differ from that 
provided by the potential (\ref{g1}). Therefore, when calculating the trajectories of infinite motion in AMD, some corrections are required to produce adequate results. In fact, such corrections are introduced in heavy-ion collision studies. As it became clear now, 
a modification of a nucleon-nucleon potential used (usually, a Volkov potential \cite{Volkov})
cannot yield correct results neither at low energies (classical trajectories do not even resemble the behavior of the quantum wave function), nor at high energies (the potential 
is deformed). Instead of such a modification, an additional two-nucleon scattering 
is introduced in AMD\cite{Onishi}. Actually, then the nucleon-nucleon scattering is taken into 
account twice; once via the Volkov potential and once via the additional scattering cross-section adjusted to the experimental value.

The minimal energy of finite motion of the particle corresponds to $R=0.$ Here the phase trajectory degenerates into the origin point. In the coordinate representation, the AMD wave function is reduced to the harmonic-oscillator ground state,
\begin{equation}
\phi_{x}^*(R=0)=\pi^{-1/4}\exp\{-{x^2\over2}\}.
\end{equation}
This approximation can be satisfactory if the value of the oscillator radius is optimal and 
the potential is deep enough. However, it looses its precision as $V_0$ decreases.

\subsection{Projection in AMD}

The projection onto the states with a definite values of the angular momentum presents a 
problem in AMD. As an example, consider the motion of a three-dimensional particle in the 
central potential field,
\begin{eqnarray}
\label{x20}
U({\bf r})=-V_0\exp\{-\alpha r^2\}.
\end{eqnarray}
The standard Hamilton function in the phase space $({\bf r},{\bf p})$ then has the form,
\begin{eqnarray}
\label{x21}
H={{\bf p}^2\over2}+U({\bf r}),~~{\bf p}^2=p_{r}^2+{M^2\over r^2},
\end{eqnarray}
where $p_r$ is the radial projection of the momentum, ${\bf M}=[{\bf
rp}]$ is the angular momentum. The projection procedure starts from fixing the value of ${\bf M}.$ 
Then a point $r=r_M$ is determined in which the function
\begin{equation}
\label{x22}
F(r,M^2)={M^2\over r^2}+U({\bf r}),
\end{equation}
reaches its minimum, and the minimal energy at given $M$,
\begin{equation}
E_0(M^2)={M^2\over2r_M^2}-V_0\exp\{-\alpha r_M^2\}
\end{equation}
is calculated. There is a maximum of $F(r,M^2)$ at $r_{\max}$; the maximum value 
$F_{\max}(M^2)$ is positive.

In the plane spanned by $(r,p_r)$, the phase trajectory shrinks to the point at the energy 
$E_0(M^2)$. If $F_{\max}(M^2)>E(M^2)>E_0(M^2),$ 
the phase trajectories are closed, concentrated around the point $r=r_M,~p_r=0.$
If $E(M^2)>F_{\max}(M^2),$ the motion becomes infinite and the phase trajectories go to the
infinity.

The Hamilton function ${\cal H}$ of the wave packet in the field of the potential (\ref{x20})
differ from (\ref{x21}) in some details only;
\begin{eqnarray}
\label{x23}
{\cal
H}={\vec{\eta}^2\over2}+{3\over4}-z^{3/2}V_0\exp\{-{1-z\over2}\xi^2\},
\end{eqnarray}
$$
\vec{\eta}^2=\eta^2_\xi+{{\bf M}^2\over\xi^2},~{\bf
M}=[\vec{\xi}\vec{\eta}].
$$
These details are the energy $3/4$ needed to create the wave packet, the factor $z^{3/2}$ 
and the factor $1-z$ instead of $\alpha$ in the exponent. There are no other differences, 
though. The minimal and, thus, the ground state, energy is reached if both $\eta_\xi$ and 
${\bf M}$ are zeros. Then the minimum corresponds to the zero value of $\xi$, 
and that gives
$$E_0={3\over4}-z^{3/2}V_0,$$
which means that a degenerate to a point trajectory exists at any small positive $V_0$.
This contradicts the statement of quantum mechanics that a bound state exists only if $V_0$
exceeds some critical value.

Furthermore, the Hamilton function (\ref{x23}) provides us with a continuous spectrum of finite
states, and some additional conditions are required to select those states which do not contradict quantum mechanical principles. The Bohr--Sommerfeld quantization rules\cite{Landau1}
or their generalization\cite{Gutz} may serve as such conditions. 

Besides, Eq.(\ref{x23}) allow for not only integer values of the angular momentum $M$, but any
other, too. In order to overcome this difficulty, the Brink--Bloch orbital has to be projected 
onto a state with the definite value of the angular momentum. This procedure is commonly known as projection before variation. Let us trace its realization on the simple example of the zero
angular momentum state. In this case, the overlap integral takes the form
\begin{eqnarray}
{1\over2}\int_{-1}^1 \exp\{RSt\}dt = {\sinh RS\over RS}.
\end{eqnarray}
The integration here is over $t$, the cosine of the angle between the vectors ${\bf R}$
and ${\bf S}$. In fact, we follow the known Peierls--Yoccoz method \cite{Peierls}. After the projection is done, the Hamilton function is
\begin{eqnarray*}
{\cal H}_0={\eta^2\over2}+{5\over4}-{\xi^2+\eta^2\over4}
\left(\coth{\xi^2+\eta^2\over2}-1\right)-
\end{eqnarray*}
\begin{eqnarray}
\label{x24}
-z^{1/2}V_0{1-\exp\{-z(\xi^2+\eta^2)\}\over{1-\exp\{-(\xi^2+\eta^2)\}}}
\exp\left(-{1-z\over2}\xi^2\right), \nonumber \\
\end{eqnarray}
where $$R=\xi+i\eta,~~S=R^*.$$
The minimization of the Hamilton function (\ref{x24}) leads to the same ground state
energy that we obtained earlier. However, the change of the potential (\ref{x20}) 
by the additional repulsive term
$$V_1\exp\{-\alpha_1r^2\},$$
quite affects the result. Now, in order to find the ground state energy, we need to minimize 
the function
$$
{5\over4}-{\xi^2\over4}
\left(\coth{\xi^2\over2}-1\right)
$$
$$
-z^{1/2}V_0{1-\exp\{-z\xi^2\}\over{1-\exp\{-\xi^2\}}}
\exp\left(-{1-z\over2}\xi^2\right)+
$$
\begin{eqnarray}
+z_1^{1/2}V_1{1-\exp\{-z_1\xi^2\}\over{1-\exp\{-\xi^2\}}}
\exp\left(-{1-z_1\over2}\xi^2\right)
\end{eqnarray}
over the semi-axis $0\leq\xi<\infty.$ If the projection before variation is not performed, 
the function to be minimized is
\begin{eqnarray}
{3\over4}-z^{3/2}V_0\exp\{-{1-z\over2}\xi^2\}+
z_1^{3/2}V_1\exp\{-{1-z_1\over2}\xi^2\}. \nonumber \\
\end{eqnarray}
In both cases, the minimum is located at different values of $\xi\ne 0$.
Evidently, the projection lowers the minimum value.

\subsection{Influence of the Pauli principle to the Hamilton function}

Having taken into account the potential (\ref{g1}) we should modify the Hamilton functions 
(\ref{u}) and (\ref{u1}). The additional term for the function of the singlet pair is
\begin{eqnarray*}
{\langle \phi_{x}^{+*}(R)|U(x)|\phi_{x}^+(R)\rangle \over
\langle \phi_{x}^{+*}(R)|\phi_{x}^+(R)\rangle }
\end{eqnarray*}
%\begin{eqnarray*}
%=-z^{1/2}V_0 \left(\exp\{-{1-z\over2}(R+S)^2\}+
%\exp\{-{1-z\over2}(R-S)^2\}\right)-
%\end{eqnarray*}
%\begin{eqnarray*}
%-z^{1/2}V_0 \tanh(RS)\left(\exp\{-{1-z\over2}(R+S)^2\}-
%\exp\{-{1-z\over2}(R-S)^2\}\right)=
%\end{eqnarray*}
\begin{eqnarray}
=-z^{1/2}V_0{\exp\{-(1-z)\xi^2\}+\exp\{-\xi^2-z\eta^2\}\over
{1+\exp\{-\xi^2-\eta^2\}}}. \nonumber \\
\end{eqnarray}
Thus, the averaging of the Gaussian potential over the symmetrized Brink--Bloch orbitals leads to
an interaction in the phase space which does not depend not only on the coordinate $\xi$ but also on the momentum $\eta$. However, at large distances between the particles, when
$\xi\gg1,$ this interaction becomes Gaussian, as it was in the previous example when the Pauli
principle didn't play any role.

As a result, the Hamiton function of the singlet pair takes the form,
\begin{eqnarray*}
{\cal H}^+(\xi,\eta)={\eta^2\over2}+{1\over4}{\xi^2+\eta^2\over4}
\left(\tanh{\xi^2+\eta^2\over2}-1\right)-
\end{eqnarray*}
\begin{eqnarray}
\label{u2}
-z^{1/2}V_0{\exp\{-(1-z)\xi^2\}+\exp\{-\xi^2-z\eta^2\}\over
{1+\exp\{-\xi^2-\eta^2\}}}.\nonumber \\
\end{eqnarray}

In the triplet pair case, the Hamilton functions is changed in a similar manner,
\begin{eqnarray*}
{\cal H}^-(\xi,\eta)={\eta^2\over2}+{1\over4}{\xi^2+\eta^2\over4}
\left(\coth{\xi^2+\eta^2\over2}-1\right)-
\end{eqnarray*}
\begin{eqnarray}
\label{u3}
-z^{1/2}V_0{\exp\{-(1-z)\xi^2\}-\exp\{-\xi^2-z\eta^2\}\over
{1-\exp\{-\xi^2-\eta^2\}}}.\nonumber \\
\end{eqnarray}

We have already stressed the dependence of the results on the depth of the potential in AMD. 
If the potential is deep both approaches (AMD and exact) show similar resluts for the ground state energy 
and the wave functions (Fig.~8). On the contrary, if the potential is so shallow that the exact solution for 
the ground-state energy appears to be just below the threshold, the AMD approach yields a positive energy
but a bound-like wave function (Fig.~9)! The same figure shows what kind of function correspond to the ground-state energy of AMD in the exact case.  
 
\begin{figure}
\centerline{\hspace{-1cm}\includegraphics[width=10.5cm]{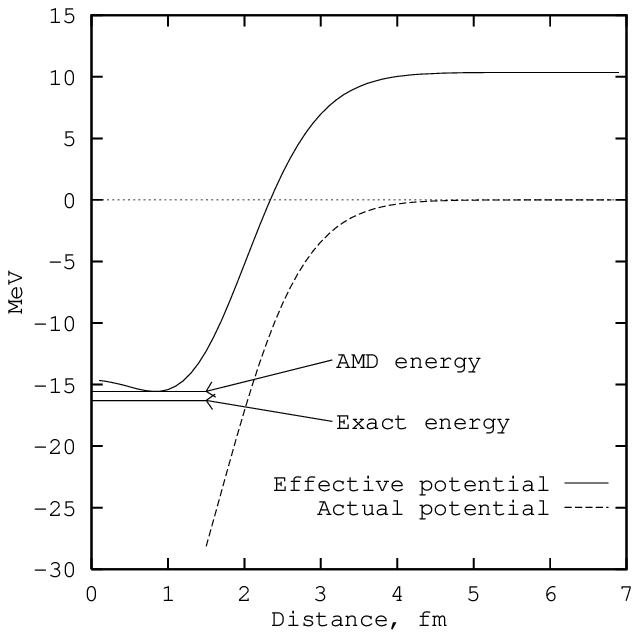}}
\centerline{\hspace{-1cm}\includegraphics[width=10.5cm]{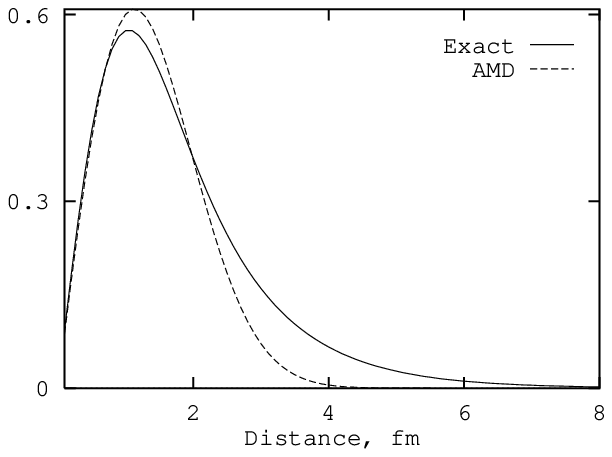}}
\caption{(Upper panel) Ground state energy of the triplet pair of fermions interacting via the deep potential 
$V(x)=-V_0 \exp(-x^2)$, $V_0=83$ MeV. The potential is shown by the solid thick line. Effective potential (see Eq.(\ref{u3})) 
for the wave packets is shown by the dashed thick line. (Lower panel) The wave functions corresponding to these energies.}
\end{figure}

\begin{figure}
\begin{center}
\centerline{\hspace{-1cm}\includegraphics[width=10.5cm]{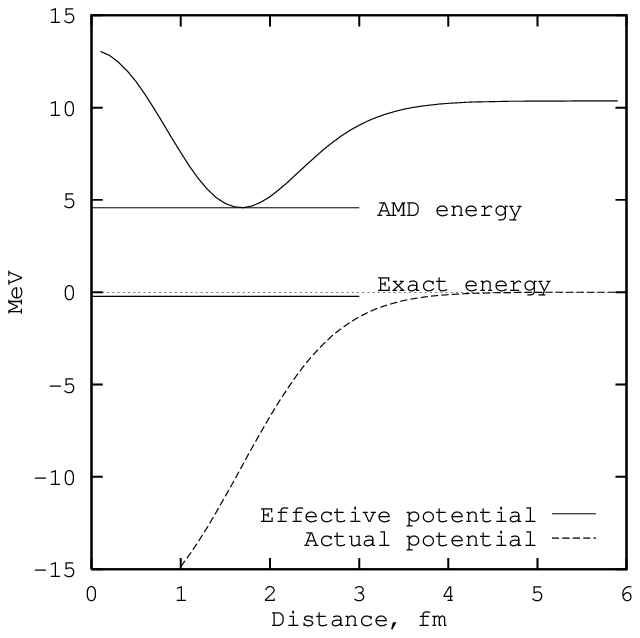}}
\centerline{\hspace{-1cm}\includegraphics[width=10.5cm]{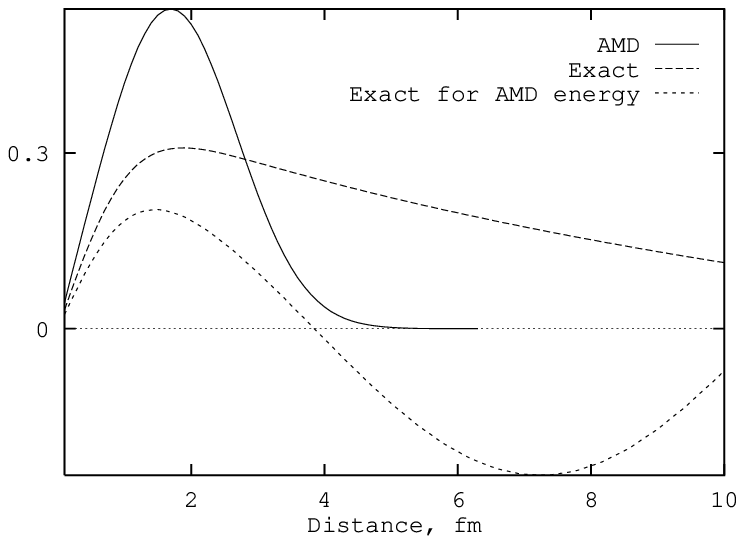}}
\end{center}
\caption{Same as in the previous figure but the potential is shallow, $V_0=33$ MeV. The short-dashed line shows the quantum-mechanical solution at $E=4.58$ MeV, belonging to the continuum.}
\end{figure}

We would like to note that the results discussed above are directly related to the variational 
calculations performed in AMD and show the way to improve them. A typical object of study for
AMD is a light nucleus featuring an $\alpha$-cluster structure. This structure is a consequence 
of the strong interaction between the nucleons constituing the same $\alpha$-cluster as compared 
to the weak interaction between the nucleons belonging to different $\alpha$-clusters, which
is due to the Pauli principle and the exchange nature of nuclear force facilitating the 
saturation of nuclear matter. We may refer to the well-known experimental fact: the binding energy of light nuclei is close to the sum of binding energies of their $\alpha$-clusters.
In AMD, the binding energy of $\alpha$-clusters is reproduced rather well but the contribution of the cluster-cluster interaction to the binding energy is underevaluated.

\subsection{AMD and RGM}

In order to explain the notion of the previos paragraph it is worthy to compare AMD to the Resonating Group Method (RGM) which also employs Brink-Bloch orbitals. Let us consider, for the 
sake of simplicity, an $\alpha$-cluster nucleus consisting of $n$ clusters ($A=4n$ nucleons). 
Whereas the AMD wave function of this nucleus is a Slater determinant of $A$ different Brink--Bloch orbitals, there are four 
times less orbitals in RGM, since four nucleons of each cluster are assumed 
to have identical orbitals. After the c.o.m. wave function is separated and a new set 
of Jacobi-like vector parameters is introduced as a linear combination of the original parameters of the Brink--Bloch orbitals, AMD provides a method to study both nucleonic and 
cluster motion. Variational AMD calculations with the Volkov potential show that the optimal values of the vector parameters for nucleons of the same cluster are identical while the vector 
parameters determining cluster locations differ from zero. The conclusion is made that there is an
$\alpha$-clusterization in light nuclei. On the other hand, this justifies the choice of 
orbitals in RGM. 

Surely, this conclusion is valid as long as the Volkov force can be considered as describing the 
real picture. For it is known that this force does not satisfy the saturation condition, 
and calculations of $sd$-shell nuclei with it does now exhibit any clusterization even if
the cluster structure is expected. 

The shape of the nucleon-nucleon function in AMD is almost independent on whether the nucleons 
it describes belong to the same cluster or not. This shape is chosen to be the best for the 
$\alpha$-cluster nucleons which leads to an error in the calculation of the cluster-cluster 
interaction energy. In RGM, the wave functions of the cluster-cluster motion is defined by an 
optimization procedure. Thus the problems facing AMD are avoided and the results are 
better\footnote{We do not discuss here new versions of AMD which essentially change the simple
physical idea of the method and, in fact, are hybrids of different aproaches\cite{Dote},\cite{Itag}.}.

In Ref.\cite{Caur}, the RGM orbitals were used in derivation and analysis of the 
classical equations of motion for the problem of scattering of the $\alpha$-particle 
off the $^{12}$C nucleus considered as three $\alpha$-particles.

It is important to note that both AMD and RGM are generalizations of the shell model, 
or rather, Elliott's SU(3) model\cite{Ell}. The latter is the limiting case in AMD and RGM,
when all vector parameters limit to zero.

\section{Spreading of wave packets}

Spreading of wave packets is a well-known phenomenon in quantum mechanics. Nevertheless, this 
phenomenon has to be neglected when the problems of the continuum are studied using dynamical 
equations of AMD. It seems to be justified if the time of interaction of colliding nucleons is
much less than the time of spreading of their wave packets. The less the energy spent to create a wave packet, the faster it spreads. For a single particle, this energy equals $3/4$. We use this simple example to show how, using classical considerations, to describe the spreading of a 
wave packet and estimate the time scale of this process.

\subsection{Wave packet of radial oscillations}

Consider the wave packet
\begin{eqnarray}
\label{i1}
\phi_r(\varepsilon)={1\over\pi^{3/4}}{1\over(1-\varepsilon)^{3/2}}
\exp\left\{-{1\over2}~{1+\varepsilon\over{1-\varepsilon}}~r^2\right\}
\end{eqnarray}
of a particle in the three-dimensional space\cite{Fil2}. It is usually used in the description of the radial oscillatory mode. The width of the packet depend on the generating parameter $\varepsilon$. The wave packet spreads apart if 
$\varepsilon$ approaches $-1$.

The overlap integral
\begin{eqnarray}
\label{k1}
\langle \phi_r(\varepsilon)|\phi_r(\varepsilon)\rangle &=&
{1\over(1-\varepsilon\varepsilon^*)^{3/2}} \nonumber \\ &=&
\sum_{n=0}^\infty {\Gamma(n+3/2)\over\Gamma(3/2)\Gamma(n+1)}
\varepsilon^n {\varepsilon^*}^n
\end{eqnarray}
provides us with an image
\begin{eqnarray}
\label{k2}
\phi_n(\varepsilon)=\sqrt{{\Gamma(n+3/2)\over\Gamma(3/2)\Gamma(n+1)}}
\varepsilon^n
\end{eqnarray}
of the basis functions of the three-dimensional harmonic oscillator with the angular momentum $L=0$, defined in the complex plane
of $\varepsilon$. That it is so is easy to see having expanded (\ref{i1}) over the powers of 
$\varepsilon$,
\begin{eqnarray}
\label{i2}
\phi_r(\varepsilon)=\sum_{n=0}^\infty \phi_n(\varepsilon)
\varepsilon^n\sqrt{{\Gamma(n+1)\over2\pi\Gamma(n+3/2)}}L^{1/2}_n(r^2)
\exp\left\{-{r^2\over2} \right\}, \nonumber \\
\end{eqnarray}
where $L^{1/2}_n(r^2)$ are the Laguerre polynomials.

The functions (\ref{k2}) are orthonormalized with the measure 
$$(1-\varepsilon\varepsilon^*)^{-1/2}{d\alpha d\beta\over\pi},~~
\varepsilon=\alpha+i\beta$$
in the unit circle $|\varepsilon|<1.$ Indeed,
\begin{eqnarray}
\int_{|\varepsilon|<1}\phi^*_{n'}(\varepsilon)\phi_n(\varepsilon)
(1-\varepsilon\varepsilon^*)^{-1/2}{d\alpha d\beta\over\pi}=\delta_{n,n'}
\end{eqnarray}

Similarly to the orbital (\ref{a1}), the wave packet (\ref{i1}) 
is an eigenfunction of the squared radius operator,
\begin{eqnarray}
\label{x11}
\hat{r^2}=-(1-\varepsilon)^2{\partial\over\partial\varepsilon}+
{3\over2}(1-\varepsilon),
\end{eqnarray}
defined in the $\varepsilon$-representation (inside the circle of unit length on the complex 
plane), i.e.,
\begin{eqnarray}
\label{x12}
\hat{r^2}\phi_r(\varepsilon)=r^2\phi_r(\varepsilon).
\end{eqnarray}

Later on we shall be considering free motion, with the Hamiltonian
\begin{eqnarray}
\hat{H}=-{1\over2}\Delta_{\bf r}.
\end{eqnarray}
The same Hamiltonian can be written in the $\varepsilon$-representation,
\begin{eqnarray}
\hat{H}={1\over2}(1+\varepsilon)^2{\partial\over\partial\varepsilon}+
{3\over2}(1+\varepsilon).
\end{eqnarray}
The eigenfunctions of the latter, corresponding to the energy
$E=k^2/2,$ are
\begin{eqnarray}
\label{x13}
\phi_k(\varepsilon)={1\over\pi^{3/4}}{1\over(1+\varepsilon)^{3/2}}
\exp\left\{-{1\over2}~{1-\varepsilon\over{1+\varepsilon}}~k^2\right\}.
\end{eqnarray}

As we had before, we now face a dilemma; either we write down the wave equation in the 
$\varepsilon$-representation and search for the exact quantum-mechanical solution to the 
problem of the radial motion, or, by studying evolution of the wave packet, we reduce the 
problem to a simpler, classical one. Beyond doubt, in the latter case we shall get only an
approximated description; a phase trajectory instead of a quantum probability distribution in the phase space. Below, we follow this way.

\subsection{Phase trajectories of radial mode}

It is easy to see that the classical Hamilton function of free motion is
\begin{eqnarray}
\label{i4}
{\cal H}={\langle \phi_r(\varepsilon)|\hat{H}|\phi_r(\varepsilon)\rangle \over
\langle \phi_r(\varepsilon)|\phi_r(\varepsilon)\rangle }=
{3\over4}{(1+\varepsilon)(1+\varepsilon^*)\over{1-\varepsilon^*\varepsilon}}. \nonumber \\
\end{eqnarray}
Eq.(\ref{i4}) is followed by the equation for the phase trajectory at a positive energy $E$,
\begin{eqnarray}
\label{y10}
(\alpha+{1\over{1+\chi}})^2+\beta^2=(1-{1\over{1+\chi}})^2,
\end{eqnarray}
where $\chi=4E/3.$ Each of the trajectories is a circle centered around the point 
$-1/(1+\chi)$ located on the real axis $\alpha$ (Fig.~10). The radius of the circle is $1-1/(1+\chi).$

\begin{figure}
\begin{center}
\includegraphics[angle=90,width=8.5cm]{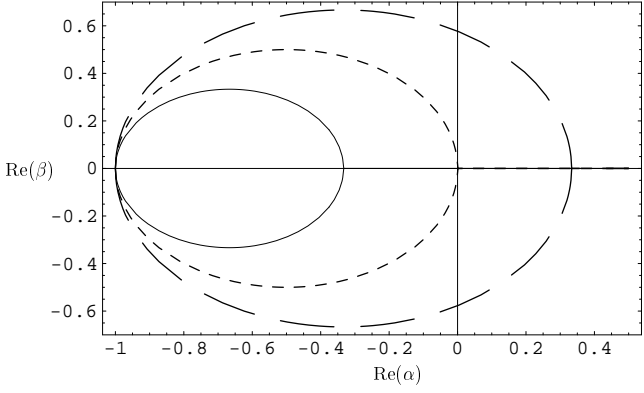}
%\epsfbox{fig12tex.eps}
\end{center}
\caption{Phase trajectories for the radial mode. Shown are the trajectories for $\chi=0.5$ ($k=\sqrt{3/4}$; solid line), 
$\chi=1.0$ ($k=\sqrt{3/2}$; short-dashed line), and $\chi=2.0$ ($k=\sqrt{3}$; solid line).}
\end{figure}

As a phase trajectory approach the point $\alpha=-1,~\beta=0,$ the width of the wave packet 
increase, and the packet completely spreads apart at this point. As mentioned above, the 
energy spent to create the packet is $E=3/4.$ In this case, 
$\chi=1,$
the radius of the phase trajectory is $1/2,$ and it joins the point 
$\varepsilon=0$ in the complex plane with the point
$\alpha=-1,~\beta=0.$

\begin{figure}
\begin{center}
\includegraphics[angle=90,width=8.5cm]{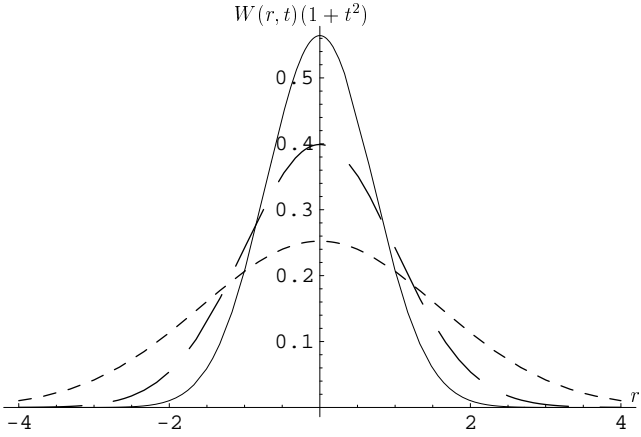}
%\epsfbox{fig13tex.eps}
\end{center}
\caption{Probability distribution function $W(r,t)$ in the coordinate space (notmalized to unity at each moment of time). 
The curves correspond to $t=0$ (solid), $t=1$ (long-dashed), and $t=2$ (short-dashed).}
\end{figure}

In order to estimate the time of spreading we turn to the classical equations of dynamics. 
These equations follow the least action principle and the expression for the Lagrange 
function\cite{Fil1},
\begin{eqnarray}
\label{i5}
{\cal L}=
{\langle \phi_r(\varepsilon)|-i{\partial/\partial t}|\phi_r(\varepsilon)\rangle -
\langle \phi_r(\varepsilon)|\hat{H}|\phi_r(\varepsilon)\rangle \over
\langle \phi_r(\varepsilon)|\phi_r(\varepsilon)\rangle }. \nonumber \\
\end{eqnarray}
As a result of the variation over $\varepsilon^*$ we arrive to the following ordinary first-order differential equation for $\varepsilon$,
\begin{eqnarray}
\label{i6}
-i\dot{\varepsilon}={1\over2}(1+\varepsilon)^2
\end{eqnarray}
or, alternatively, to the set of equations for $\alpha$ and $\beta$,
\begin{eqnarray}
\dot{\alpha}=-(1+\alpha)\beta,~~
\dot{\beta}={1\over2}\{(1+\alpha)^2-\beta ^2\}.
\end{eqnarray}
Integration of Eq.(\ref{i6}) with the initial condition 
$\varepsilon(0)=0$ yields
\begin{eqnarray}
\label{j1}
\alpha(t)=-{t^2\over{4+t^2}},~~\beta(t)=-{2t\over{4+t^2}}.
\end{eqnarray}
The function $\beta(t)$ reaches its minimum at $t=2,$ when the phase trajectory has elapsed the 
first quarter of the circle. It takes infinite amount of time to elapse the second quarter.

Having used Eqs. (\ref{j1}),(\ref{i1}), and taking into account the normalization factor, we 
obtain the time dependence of the probability density of values of $r$ in the coordinate 
representation.
\begin{eqnarray}
\label{j2}
W(r,t) &\equiv&
{\phi_r(\varepsilon)\phi_r(\varepsilon)\over
\langle \phi_r(\varepsilon)|\phi_r(\varepsilon)\rangle } \nonumber \\ &=&
{1\over\pi^{3/2}}~{1\over(1+t^2)^{3/2}}\exp\left(-{r^2\over{1+t^2}}\right).
\end{eqnarray}
Thus, the time dependence of a wave packet at rest is explicitly found. It is 
illustrated by Eq.~(\ref{j2}) and Fig.~11.

Finally, in order to understand the relation between the quantum result in the Fock--Bargmann
space and the classical result, it is useful to compare the phase trajectory (\ref{y10}) 
with the quantum distribution in the phase plane for the state $\phi_k(\varepsilon)$ 
with the kinetic energy $k^2/2.$
\begin{eqnarray*}
\phi^*_k(\varepsilon)\phi_k(\varepsilon)(1-\varepsilon\varepsilon^*)^{-1/2}=
\pi^{-3/2}\left({1\over(1+\varepsilon)(1+\varepsilon^*)}\right)^{3/2}
\end{eqnarray*}
\begin{eqnarray}
\label{y11}
\times \exp\{-{1-\varepsilon\varepsilon^*\over(1+\varepsilon)(1+\varepsilon^*)}k^2\}
(1-\varepsilon\varepsilon^*)^{-1/2}.
\end{eqnarray}

\subsection{Spreading of Brink--Bloch wave packet}

Finally we show how to modify the expression for the Brink--Bloch wave packet in order to 
completely take into account its evolution due to its spreading as it moves in the three-dimensional space. Assume
\begin{eqnarray*}
\phi_{\bf r}(\varepsilon,{\bf R})={1\over\pi^{3/4}}
{1\over(1-\varepsilon)^{3/2}}
\exp\{-{1\over2}~{1+\varepsilon\over{1-\varepsilon}}~r^2
\end{eqnarray*}
\begin{eqnarray}
\label{i7}
+{\sqrt{2}\over{1-\epsilon}}({\bf Rr})-
{1\over2}~{1-2\varepsilon^*+\varepsilon\varepsilon^*\over{1-\varepsilon}}
~{R^2\over{1-\varepsilon\varepsilon^*}}\}.
\end{eqnarray}
The overlap integral of (\ref{i7}) is
$$
\langle \phi_{\bf r}(\varepsilon,{\bf R})|\phi_{\bf r}(\varepsilon,{\bf R})\rangle =
{1\over(1-\varepsilon\varepsilon^*)^{3/2}}$$
\begin{eqnarray}
\label{i8}
\times \exp\left\{{({\bf RS})+\varepsilon^*R^2/2+\varepsilon S^2/2\over
{1-\varepsilon\varepsilon^*}}\right\}.
\end{eqnarray}
The kinetic energy of the wave packet now takes the form
\begin{eqnarray*}
{\langle \phi_{\bf r}(\varepsilon,{\bf R})|\hat{T}|\phi_{\bf r}(\varepsilon,{\bf
R})\rangle 
\over\langle \phi_{\bf r}(\varepsilon,{\bf R})|\phi_{\bf r}(\varepsilon,{\bf R})\rangle }=
\{{3\over4}~{(1+\varepsilon)(1+\varepsilon^*)
\over{1-\varepsilon\varepsilon^*}}-
\end{eqnarray*}
\begin{eqnarray}
\label{i9}
-{1\over4}~{[(1+\varepsilon^*){\bf R}-(1+\varepsilon){\bf S}]^2\over
(1-\varepsilon\varepsilon^*)^2}\}
\langle \phi_{\bf r}(\varepsilon,{\bf R})|\phi_{\bf r}(\varepsilon,{\bf R})\rangle . \nonumber \\
\end{eqnarray}
Finally, the potential energy is
\begin{eqnarray*}
{\langle \phi_{\bf r}(\varepsilon,{\bf R})|-V_0\exp\{-(z-1)/z~r^2\}|
\phi_{\bf r}(\varepsilon,{\bf R})\rangle \over
\langle \phi_{\bf r}(\varepsilon,{\bf R})|\phi_{\bf r}(\varepsilon,{\bf R})\rangle }
\end{eqnarray*}
\begin{eqnarray}
=-z^{3/2}V_0\left({\Delta\over\Delta_z}\right)^{3/2}
\end{eqnarray}
$$
\times \exp\left\{-{(1-z)\over2}~{\Delta\over\Delta_z}
\left({(1-\varepsilon^*){\bf R}+(1-\varepsilon){\bf S}
\over{1-\varepsilon\varepsilon^*}}\right)^2\right\},
$$
where $$\Delta=1-\varepsilon\varepsilon^*,~~
\Delta_z=
z(1-\varepsilon\varepsilon^*)+(1-z)(1-\varepsilon)(1-\varepsilon^*).$$

These expressions suffice to derive the classical equations of dynamics and study the problem 
of spreading of the Brink--Bloch wave packet and its influence to motion of the packet in the 
field of a Gaussian potential.

The integrals of free motion are kinetic energy (equal to the total energy $E$)
\begin{eqnarray}
\label{i10}
{3\over4}~{(1+\varepsilon)(1+\varepsilon^*)
\over{1-\varepsilon\varepsilon^*}}-
{1\over4}~{[(1+\varepsilon^*){\bf R}-(1+\varepsilon){\bf S}]^2\over
(1-\varepsilon\varepsilon^*)^2}=E, \nonumber \\
\end{eqnarray}
and the momentum
\begin{eqnarray}
\label{i11}
-{i\over\sqrt{2}}~{(1+\varepsilon^*){\bf R}-(1+\varepsilon){\bf S}\over
1-\varepsilon\varepsilon^*}={\bf P}.
\end{eqnarray}
Eqs.~(\ref{i10}) and (\ref{i11}) are followed by the relation
\begin{eqnarray}
\label{i12}
{3\over4}~{(1+\varepsilon)(1+\varepsilon^*)
\over{1-\varepsilon\varepsilon^*}}=E-{1\over2}{\bf P}^2.
\end{eqnarray}

Thus, we reformulated the problem in such a way that the continuum begins at zero energy
and there is no need to apply any special technique to suppress the packet creation energy at
large distances. As a result, still remaining in the framework of AMD but having redefined 
the Brink--Bloch orbitals by introducing the degrees of freedom describing the wave packet spreading, we can consider not only scattering but also resonance states in the relatively low energy region.

\section{Conclusion}

Trying to describe collisions of nuclei as collisions of their nucleons by the classical 
equations of dynamics we face a number of problems. Let us summarize them.

i) The Pauli principle affects the derivation of the classical equations in the region of 
low values of coordinates and momenta where quantum corrections are expected to be significant.
However, the corrections due to the Pauli principle do not suffice to describe some important 
features of wave functions by phase trajectories if the energy of relative motion of the 
nucleons is low. At the same time, as energy increases, the classical results limit to the 
quantum ones. 

ii) The kinetic energy of classical motion of nucleons contains constant factors which raise 
the threshold energy of break-up channels as well as the total break-up energy. This 
unfortunate phenomenon can be eliminated by introducing those degrees of freedom which 
describe the spreading of wave packets. The spreading affects the process as long as the 
packets move slowly and their width can change within the time of the nuclear reaction.

iii) In AMD, the nucleon-nucleon interaction is taken into account twice. First, it forms 
the trajectory of each nucleon. Second, it is responsible for the effective cross-section of the
nucleon-nucleon collision. A similar situation occurs when the Vlasov equation is used for the 
description of a system of particles if, along with the mean-field potential, a collision term 
is introduced. This term should be then expressed via the scattering cross-sections formed by
the residual interaction. It is understood that the Vlasov (or Boltzmann) equation is applicable 
only if the nucleon density is low. Therefore their usage in studies of heavy-ion collisions 
require additional justification.

iv) In the ground states of even-even nuclei, the AMD wave function features an $\alpha$-cluster
structure or, if the number of neutrons exceeds the number of protons, some of the clusters 
are singlet pairs of neutrons. It is the Volkov force which leads to this conclusion. Therefore,
to describe collisions of such nuclei, effective cross-sections of elastic and inelastic scattering of $\alpha$- and di-neutron clusters should be used. Although there are experimental 
data for the $\alpha-\alpha$ scattering available, the cross-sections of scattering of singlet 
di-neutrons have to be calculated theoretically. The nucleons themselves appear after the first
collision only, with some probability. Besides, inelastic scattering may produce deutron (d), triton (t) and helionic (h) clusters.  Later they all interact with alphas and di-neutrons. This
leads to the stochastisation, and the process ceases to be deterministic.

%\newpage


\begin{thebibliography} {99}

\bibitem{exp1} B. Jakobsson, G.J\"{o}nson, B. Lindquist and A.Oskarsson,
Z. Phys. {\bf A307}, 293 (1982); Nucl. Phys. {\bf A509}, 195 (1990).\\
        H. Gutbrod et al., Phys. Lett. B {\bf 216}, 267 (1989); Phys. Rev. C.
{\bf 42}, 640 (1990).\\
        K. Hagel et al.,~Phys.~Rev.~C. {\bf 50}, 2017 (1994).

\bibitem{theor1} J.~Aichelin and H.~St\"{o}cker, Phys.~Lett.
{\bf B176}, 14 (1986).\\
        J.~Aichelin, Phys.~Rep. {\bf 202}, 235 (1991).

\bibitem{Hor1} A.~Ono, H.~Horiuchi, T.~Maruyama and A.~Ohnishi, Phys.~Rev.
~Lett. {\bf 68}, 2898 (1992) \\
        T.~Maruyama, A.~Ono, A.~Ohnishi, and H.~Horiuchi,
Progr.~Theor.~Phys.
{\bf 87}, 1367 (1992). \\
        A.~Ono, H.~Horiuchi, and T.~Maruyama,~Phys.~Rev.~C. {\bf 48}, 2946
(1993). \\
        A.~Ono, H.~Horiuchi,~Phys.~Rev.~C. {\bf 51}, 299
(1995). \\
        Y. Kanada-En'yo, H.~Horiuchi, and A.~Ono,~Phys.~Rev.~C. {\bf 52},
628
(1996). \\
        A.~Ono, H.~Horiuchi,~Phys.~Rev.~C. {\bf 53}, 844
(1996). \\
        A.~Ono, H.~Horiuchi,~Phys.~Rev.~C. {\bf 53}, 2958
(1996).

\bibitem{Fil1} G.~F.~Filippov, S.~V.~Mokhov, A.~M.~Sycheva, K.~Kat\=o, S.~V.~Korennov,
Phys.~Atom.~Nucl. {\bf 62}, 95 (1999).

\bibitem{F-B} V.~A.~Fock. Z.~Phys. {\bf 49}, 339 (1928). \\
V.~Bargmann, Rev.~Mod.~Phys. {\bf 34}, 829 (1962).

%\bibitem{Per} A.~M.~Perelomov, Generalized coherent states and their application, 
%M., Nauka, 1987.

\bibitem{Feyn-Wign} R.~Feynman, Statistical Mechanics. (Springer-Verlag, 1986).

\bibitem{Br} D.~M.~Brink, in: International School of Physics "Enrico
Fermi",
Course 37, Varenna, Italy (Academic, N.~Y., 1965).

\bibitem{Barg} V.~Bargmann, Comm.~Pure~Appl.~Math. {\bf 14}, 187 (1961).

\bibitem{Landau1} L.~D.~Landau, E.~M.~Lifshitz, Quantum Mechanics: non-relativistic Theory. (Pergamon Press, 1977)

\bibitem{Landau2} L.~D.~Landau, E.~M.~Lifshitz, Statistical Physics (Pergamon Press, 1980)

\bibitem{Fil3} G.~F.~Filippov, Phys.~Atom.~Nucl. {\bf 58}, 1963 (1995).

\bibitem{Ell} J.P.Elliott, Proc. Roy. Soc. A. {\bf 245}, 128 (1958).

\bibitem{Bat} G.~Bateman, A.~Erdelyi, Higher Transcedental Functions, vol.2 
(Krieger Pub., 1981).

\bibitem{Fil2} G.F.Filippov, Riv. Nuovo Cim. {\bf 9}, 1 (1989).

\bibitem{Volkov} A.Volkov, Nucl. Phys. {\bf 75}, 33 (1965).

\bibitem{Onishi} A.~Ono, H.~Horiuchi, T.~Maruyama and A.~Ohnishi,
Progr.~Theor.~Phys.
{\bf 87}, 1185 (1992).


\bibitem{Gutz} M.~C.~Gutzwiller, J. Math. Phys., 1971, {\bf 12}, 343.

\bibitem{Peierls} R.~E.~Peierls, J.~Yoccoz, Proc. Roy. Soc. London A, 1957,
{\bf 70}, 3811.

\bibitem{Dote} A.~Dot\'{e} and H.~Horiuchi, Progr.~Theor.~Phys.
{\bf 103}, 91 (2000). \\
        A.~Dot\'{e} and H.~Horiuchi, Progr.~Theor.~Phys.
{\bf 103}, 261 (2000).

\bibitem{Itag} N.~Itagaki and S.~Aoyama,~Phys.~Rev.~C. {\bf 61}, 024303-1
(2000).

\bibitem{Caur} E.~Caurier, B.~Grammaticos and T.~Sami, Phys. Lett., 1982,
{\bf B109}, 150.


\end{thebibliography}
\end{document}